\documentclass[12pt]{iopart}
\usepackage{graphicx}
\usepackage{color}
\begin{document}

\title{Dynamics of a single exciton in strongly correlated bilayers}
\author{Louk Rademaker, Kai Wu and Jan Zaanen}
\address{Institute-Lorentz for Theoretical Physics, Leiden University, PO Box 9506, NL-2300 RA Leiden, The Netherlands}
\ead{rademaker@lorentz.leidenuniv.nl}

\begin{abstract}
We formulated an effective theory for a single interlayer exciton in a bilayer quantum antiferromagnet, in the limit that the holon and doublon are strongly bound onto one interlayer rung by the Coulomb force. Upon using a rung linear spin wave approximation of the bilayer Heisenberg model, we calculated the spectral function of the exciton for a wide range of the interlayer Heisenberg coupling $\alpha=J_{\perp}/Jz$. In the disordered phase at large $\alpha$, a coherent quasiparticle peak appears representing free motion of the exciton in a spin singlet background. In the N\'{e}el phase, which applies to more realistic model parameters, a ladder spectrum arises due to Ising confinement of the exciton. The exciton spectrum is visible in measurements of the dielectric function, such as $c$-axis optical conductivity measurements.
\end{abstract}


\submitto{\NJP}

\maketitle
\section{Introduction}
An exciton is the bound state of an electron and a hole, and considering their bosonic character the question immediately arises whether they can condense into an exciton Bose condensate\cite{Moskalenko:2000p4767}. The quest for such exciton superfluidity has, over the past decade, increasingly focussed its attention to layered structures where one layer contains holes and the other layer contains electrons\cite{Butov:2007fr}. The Coulomb attraction between the electrons and holes then allows for the formation of so-called interlayer excitons. In 2004 a condensate of interlayer excitons was successfully created in a heterostructure of two 2DEGs under the application of a perpendicular magnetic field\cite{Eisenstein:2004go}. Since then many other candidate materials were suggested that should support interlayer exciton condensation in the absence of magnetic fields, such as graphene\cite{Dillenschneider:2008dp,Zhang:2008kh,Min:2008id,Lozovik:2008ug} or topological insulators\cite{Seradjeh:2009p4980}. One class of candidate materials has not been considered yet, namely the Mott insulators\cite{Imada:1998p2790}. The strong interactions between electrons make these materials currently one of the most fascinating and the least understood solid state compounds. When making heterostructures of $p$ and $n$-doped quasi-two-dimensional CuO$_2$ layers one expects the formation of interlayer excitons, and these excitons will interact strongly with magnetic excitations, possibly leading to unexpected dynamics. To explore all these unexpected dynamics of the excitons in the strongly correlated system such as the exciton condensation, understanding the dynamics of single exciton will be the first step.

Heterostructures of $p$ and $n$-doped cuprates can be typically described by a strongly correlated model: the bilayer $t-J$ model, which is extended from two single-band $t-J$ models for each layer with coupling terms between the layers as following:
\begin{equation}
  H_{bt-J}=H_t+H_J+H_V\label{btJ}
\end{equation}
where $H_t$ is the hopping of electrons in each layer
\begin{equation}
  H_{t}=-t_e\sum_{\langle ij \rangle\sigma,l}c_{il\sigma}^\dagger c_{jl\sigma}+h.c.\label{Ht}
\end{equation}
and $H_J$ is the bilayer Heisenberg model describing the undoped Mott insulating state 
\begin{equation}
  H_J=J\sum_{\langle ij \rangle,l}\mathbf{s}_{il}\cdot\mathbf{s}_{jl}+J_{\perp}\sum_{i}\mathbf{s}_{i1}\cdot\mathbf{s}_{i2}. \label{HJ}
\end{equation}
Here $c_{il\sigma}$ and $\mathbf{s}_{i l}$ denotes the electron and spin operators respectively on site $i$ in layer $l=1,2$. The Heisenberg $H_J$ is antiferromagnetic with $J>0$ and $J_{\perp}>0$.
The last term $H_V$ in (\ref{btJ}) is the Coulomb attraction between a vacant site (holon) and double-occupied site (doublon) in the same rung, described by
\begin{equation}
  H_V= V\sum_{i}n_{i1}n_{i2}\label{HV}
\end{equation}
which is the force required to form an exciton in the same rung. Without loss of generality, we assume that layer '1' contains the excess electrons with the constraint $\sum_{\sigma}c_{i1\sigma}^\dagger c_{i1\sigma}\geq1$ and layer '2' has the constraint $\sum_{\sigma}c_{i2\sigma}^\dagger c_{i2\sigma}\leq1$. If one considers doped systems, this amounts to $n$-type doping in layer '1' and $p$-type in layer '2'.

\begin{figure}
 \includegraphics[width=8.6cm]{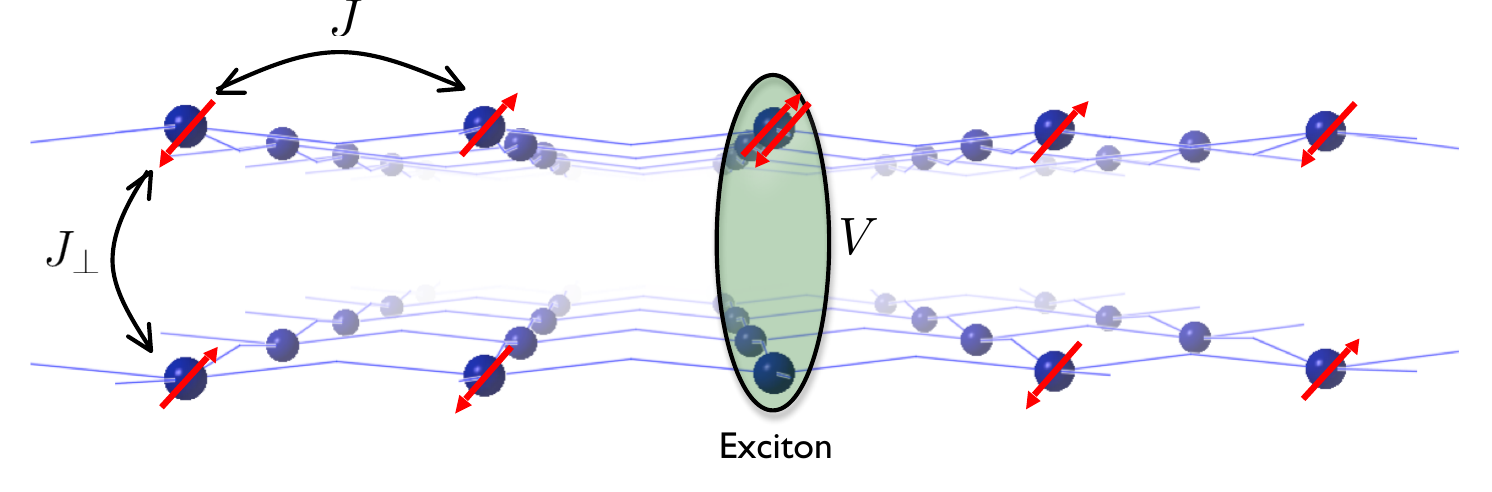}
 \caption{\label{FigBExciton}Naive real space picture of an exciton in a strongly correlated material, as viewed from the side. Two square lattices (blue balls) are placed on top of each other. The red arrows denote the spin ordering, which forms a perfect N\'{e}el state. The exciton consists of a bound pair of a double occupied and a vacant site on an interlayer rung. The energy required to break this doublon-holon pair is $V$. The magnetic ordering is governed by the in-plane Heisenberg $J$ and the interlayer $J_\perp$, as described by the Hamiltonian (\ref{HJ}).}
\end{figure}

In this paper we will present a theoretical framework describing the dynamical properties of a single exciton in a strongly correlated bilayer described by (\ref{btJ}), following our previous shorter publication on this topic\cite{Rademaker2011EPL}. The binding of the holon and the doublon is determined by the interlayer Coulomb repulsion and we will focus on the strong coupling limit $(V>t)$. This implies that the exciton is formed by the holon and doublon on the same rung, as is shown in Figure \ref{FigBExciton}.

\begin{figure}
 \includegraphics[width=8.6cm]{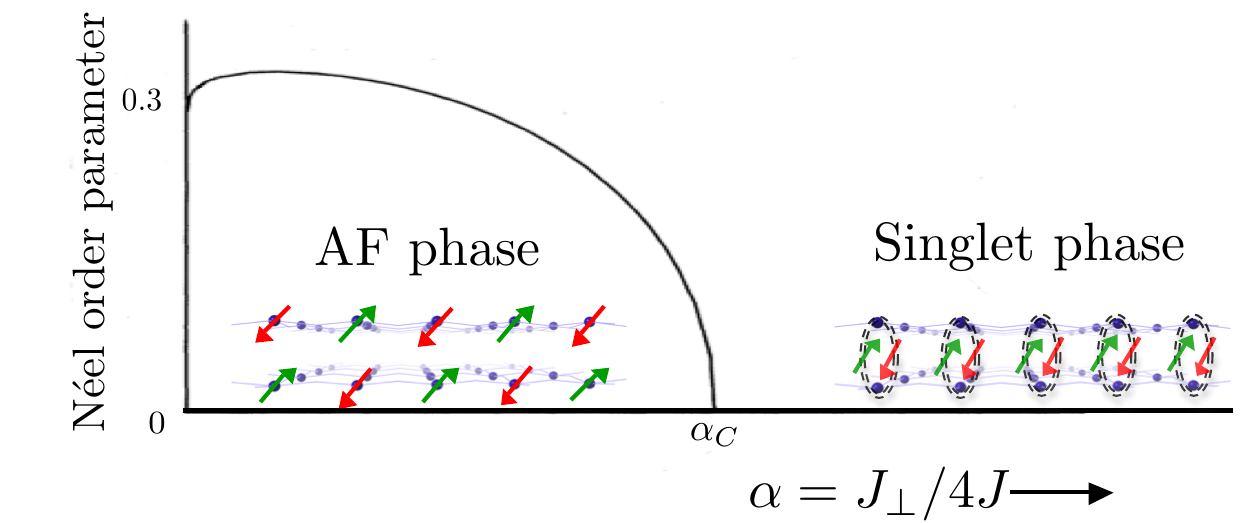}
 \caption{\label{FigHeisenbergPD}Zero temperature phase diagram of the bilayer Heisenberg model as a function of interlayer coupling strength $\alpha = \frac{J_\perp}{4J}$ on the horizontal axis. At a critical value $\alpha_c$ a quantum phase transition exists from the antiferromagnetic to the singlet phase. The vertical axis shows the N\'{e}el order parameter signaling antiferromagnetism. Note that even at $\alpha = 0$ the N\'{e}el order parameter is reduced from the mean field value $\frac{1}{2}$ to approximately $0.3$ due to spin flip interactions. (Adapted from Ref. \cite{Chubukov:1995p2296}.)}
\end{figure}

Understanding of the bilayer Heisenberg model will be an important step towards analysing the dynamics of a single exciton. The ground state and excitations of the bilayer Heisenberg Hamiltonian have been studied quite extensively using Quantum Monte Carlo (QMC) methods\cite{Sandvik:1995p5136,Sandvik:1994p3011}, dimer expansions\cite{Weihong:1997p5123,Gelfand:1996p5121,Hida:1992p5117} and the closely related bond operator theory\cite{Matsushita:1999p5145,Yu:1999p5144}, the nonlinear sigma model\cite{Duin:1997p2301,Chakravarty:1989p1442} and spin wave theory\cite{Miyazaki:1996p5122,Millis:1993p5137,Matsuda:1990p5116,Hida:1990p5115}. All results indicate a O(3) universality class quantum phase transition at a critical value of $J_\perp / J$ from an antiferromagnetically ordered to a disordered state, see Figure \ref{FigHeisenbergPD}. A naive mean field picture of the antiferromagnetic ground state is provided by the N\'{e}el state, in which each of the sublattices are occupied by either spin up or spin down electrons as shown in Figure \ref{FigBExciton}. However, the exact ground state is scrambled up by spin flip interactions reducing the N\'{e}el order parameter to about 60\% of its mean field value\cite{Manousakis:1991p2291}. A finite interlayer coupling $J_\perp$ influences the antiferromagnetic order. In the limit of infinite $J_\perp$, the electrons on each interlayer rung tend to form singlets destroying the antiferromagnetic order.

Standard spin wave theories however cannot account for the critical value of $J_\perp / J \sim 2.5$ found in QMC and series expansion studies. This discrepancy between numerical results and the spin wave theory has a physical origin. Chubukov and Morr\cite{Chubukov:1995p2296} pointed out that standard spin wave theories do not take into account the longitudinal (that is, the interlayer) spin modes. By taking into account those longitudinal spin waves one can derive analytically the right phase diagram\cite{Sommer:2001p5326}. Another correct method is to introduce an auxiliary interaction which takes care of the hard-core constraint on the spin modes\cite{Kotov:1998p5125}.

\begin{figure}
 \includegraphics[width=8.6cm]{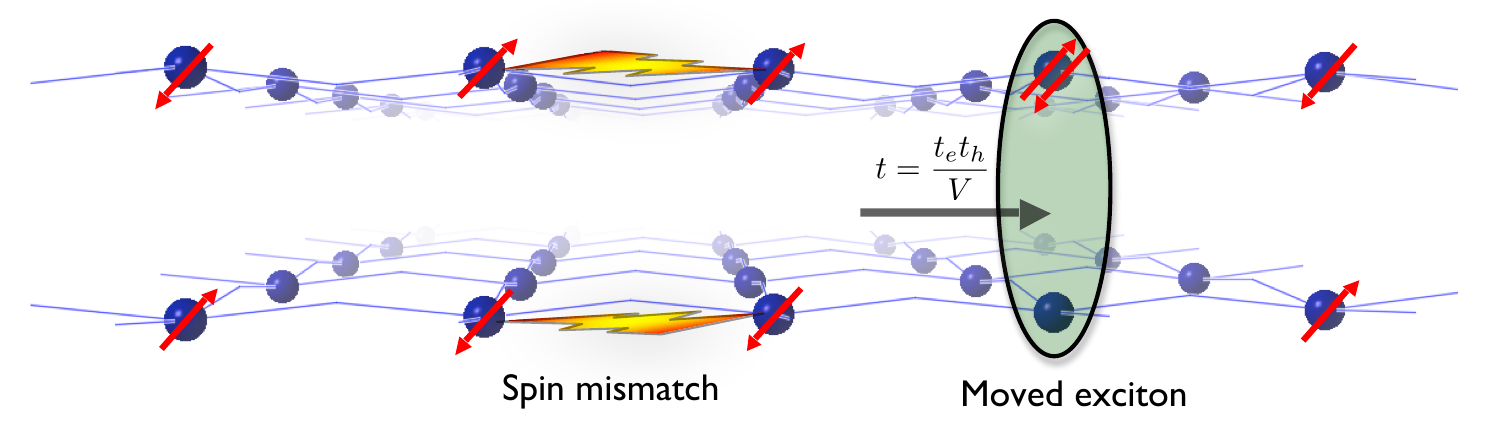}
 \caption{\label{FigBExcitonDynamic}Exciton motion in a naive real space picture. In a perfect N\'{e}el state, the motion of an exciton (with respect to the situation in Figure \ref{FigBExciton}) causes a mismatch in the spin ordering. The kinetic energy gained by moving the exciton is proportional to the energies of the doublon $t_e$ and holon $t_h$ divided by the exciton binding energy $V$.}
\end{figure}

If one wants to study the doped bilayer antiferromagnet however, one needs explicit expressions of how a moving dopant (be it a hole, electron or exciton) interacts with the spin excitations. Even though the N\'{e}el state is just an approximation to the antiferromagnetic ground state, it provides an intuitive explanation of the major role spins play in the dynamics of any dopant. As can be seen in Figure \ref{FigBExcitonDynamic}, a moving exciton causes a mismatch in the previously perfect N\'{e}el state. Consequently, the motion of an exciton is greatly hindered and a full understanding of possible spin wave interactions is needed to describe the exciton dynamics. This is of course similar to the motion of a single hole in a single Mott insulator layer\cite{SchmittRink:1988p10,Kane:1989p585}. Vojta and Becker\cite{Vojta:1999p5138} have computed the spectral function of a single hole in the Heisenberg bilayer. A rung linear spin wave approximation\cite{Sommer:2001p5326} is needed to obtain the expressions for the spin waves in terms of single site spin operators. Summarizing, we will formulate first an effective \emph{exciton $t-J$ model} from the bilayer $t-J$ model in the limit of strong Coulomb attraction in section \ref{ExcitonParagraph}. In order to find the interaction coefficients between excitons and spin excitations we will construct a spin wave theory of the bilayer Heisenberg model in section \ref{SectionHeisenberg}. Based on these two developments, we can compute the exciton spectral function using the self-consistent Born approximation in section \ref{SectionSCBA}. Finally, we connect the exciton spectral function to measurable quantities in section \ref{SectionExperiment}.

\section{The bilayer exciton $t-J$ model}
\label{ExcitonParagraph}
The bilayer $t-J$ model (\ref{btJ}) describes generally the $p$/$n$-doped bilayer antiferromagnet. The behavior of a bound exciton however depends on the magnitude of the Coulomb force $V$ in $H_V$, equation (\ref{HV}). If the Coulomb repulsion is relatively weak, the motion of holons and doublons will be relatively independent with each other and the $H_V$ can be treated as a perturbation on $H_t+H_J$. The full exciton-susceptibility $\chi(\omega)$ can be obtained from the bare susceptibility $\chi_0(\omega)$ in the absence of the Coulomb force using the ladder diagram approximation,
\begin{equation}
	\chi(\omega) = \frac{\chi_0(\omega)}{1 - V \chi_0(\omega)}.
	\label{LadderApp}
\end{equation}
Since the undoped state is a Mott insulator, there is a gap in the imaginary part of the bare susceptibility $\chi_0''$. Above this gap there is an onset of the particle-hole continuum. In the ladder diagram approximation, there can only be a single delta function peak in the full susceptibility at $V \chi_0' = 1$ signaling the formation of an exciton. We conclude that in the weak coupling limit no special exciton features other than a single delta function peak can appear in the gap. Following our expectation that realistic materials are in fact in the strong coupling limit, as explained in section \ref{SectionExperiment}, we will henceforth focus our attention to the strong coupling limit.

\begin{figure}
 \includegraphics[width=8.6cm]{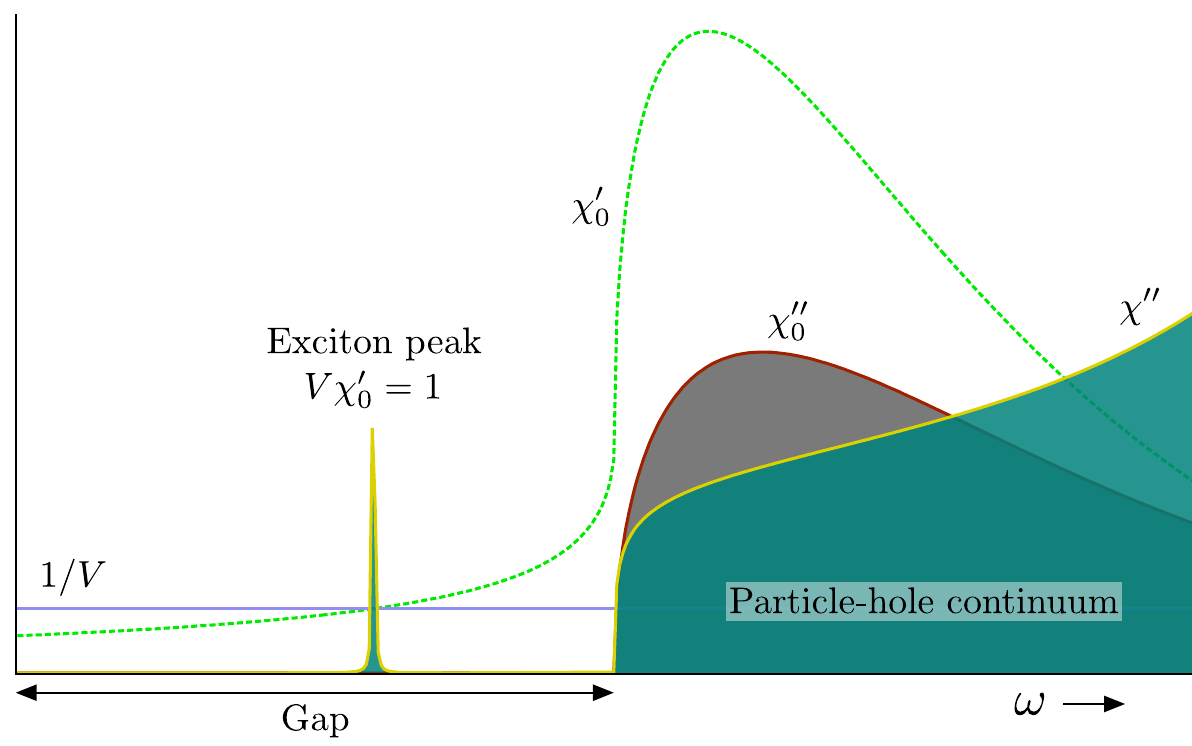}
 \caption{\label{susceptibility2}In weak coupling the spectrum of an exciton is obtained by the ladder diagram approximation from the spectrum of the single doped hole. The $\chi''_0$ and $\chi'_0$ are respectively the imaginary and real part of the bare exciton susceptibility. The $\chi''$ is the imaginary part of the full exciton susceptibility obtained in the ladder diagram approximation (\ref{LadderApp}). Besides the continuous particle-hole spectrum above the gap, there can only be a single exciton peak determined by $V\chi'_0=1$ in the weak coupling limit.}
 \end{figure}

In the strongly coupling limit $(V\gg t)$, the hopping term $H_t$ can be treated as a perturbation on the unperturbed $H_V$ using the perturbation method developed by Kato\cite{Kato:1949p3123}, in a manner similar to the derivation of the $t-J$ model from the Hubbard model\cite{Klein:1973p3139,Takahashi:1977p3119,Chao:1977p5127}. In this method, one considers first an exact solvable part of the Hamiltonian, in this case the interlayer Coulomb interaction $H_V$. It has the eigenvalues
\begin{equation}
	E_{\widetilde{N}} = V ( N - N_0 + \widetilde{N} ) = E_0 + V \widetilde{N}
\end{equation}
where $N$ is the total number of sites, $N_0$ is the number of dopants per layer and $\widetilde{N}$ is the number of double occupied sites that do not lie above a vacant site. It is clear that the ground state of $H_V$ is given by the state where all double occupied and vacant sites lie above each other, as depicted in Figure \ref{FigBExciton}. As mentioned before an \emph{exciton} consists of a double occupied and a vacant site bound on top of each other. Consequently, the ground state of $H_V$ is the state where all dopants are bound into excitons.

\begin{figure}
 \includegraphics[width=8.6cm]{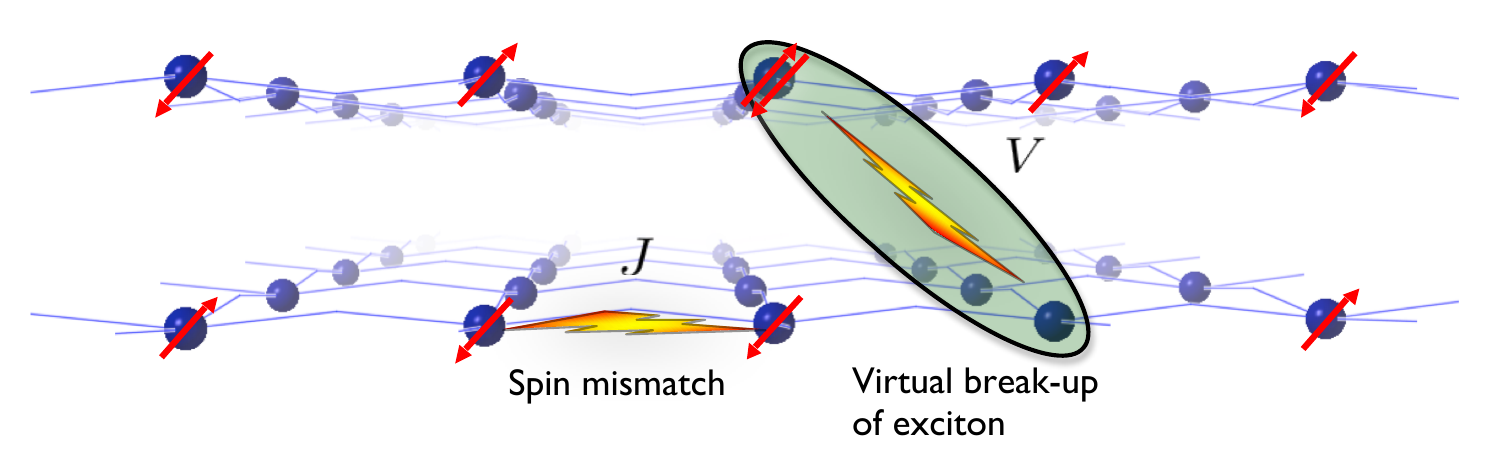}
 \caption{\label{FigExciton}The motion of the composite exciton can be related to the motion of its constituents via Kato's perturbation method. In this method a virtual intermediate breakup of the exciton is in between the initial state (Figure \ref{FigBExciton}) and the final state (Figure \ref{FigBExcitonDynamic}). The kinetic energy of the exciton is therefore the product of the kinetic energies of the holon and doublon divided by the energy of this virtual state, $t_{ex} = t_e t_h /V$.}
\end{figure}

The essence of Kato's perturbation method is that we now forbid all states with higher $H_V$ eigenvalues. In our model, this implies that we forbid states such as the one depicted in Figure \ref{FigExciton} where the double occupied site is not on top of the vacant site. In zeroeth order, hopping of electrons is forbidden since that would break up an exciton state. Therefore the zeroeth order Hamiltonian only contains Heisenberg terms $H^{(0)} =H_J$.

In second order processes are allowed that virtually break up excitons, but end up with only bound excitons. The corresponding effective Hamiltonian is given by
\begin{equation}
	-\frac{1}{2V} P_e \left( H_t \right) (1 - P_e)
	\left( H_t\right) P_e
\end{equation}
where $P_e$ is the operator that projects out states with unbound dopants. As can be verified from Figure \ref{FigExciton} this process allows the hopping of excitons by virtually breaking the dopants apart. If we define the exciton operator in terms of electron creation operators
\begin{equation}
	E^\dagger_i =
		c_{i1 \uparrow}^\dagger c_{i1 \downarrow}^\dagger
		( 1 - \rho_{i2} ),
\end{equation}
where $\rho_{i2} = \sum_\sigma c^\dagger_{i2\sigma} c_{i2\sigma}$ is the density operator in the $p$-type layer. The exciton hopping process can be formulated as
\begin{equation}
	H_{t,ex}
		= - \frac{t_et_h}{V} \sum_{<ij>\sigma \sigma'}
		E^\dagger_j \left[ c^\dagger_{i1 \sigma'} c^\dagger_{i 2 \sigma}
		c_{j 2 \sigma} c_{j 1 \sigma'} \right] E_i
	\label{ExcitonHopping}
\end{equation}
Note that in this Hamiltonian, no break-up of the exciton is required. The virtual process as described before only enabled us to relate the single layer kinetic energies to the bilayer exciton kinetic energy,
\begin{equation}
	t = \frac{t_e t_h}{V}.
\end{equation}
Here $t_e$ is the hopping energy for a single electron, $t_h$ the hopping energy for a single hole and $t$ is the hopping energy for a bound exciton. In addition to this hopping process there are also second order processes that equal a shift in chemical potential of the excitons. In the limit that we are interested in, that of a single exciton, we neglect chemical potential terms.

In conclusion, we formulated a model for the strong coupling limit of $H_V$ that describes the motion of bound excitons in a Mott insulator double layer. The corresponding Hamiltonian is
\begin{equation}
	H = H_{t,ex} +H_J
	\label{Excitont-J1}
\end{equation}
We will refer to this model as the \emph{exciton $t-J$ model}.

\subsection{The singlet-triplet basis}
The hopping term (\ref{ExcitonHopping}) represents an exciton $E_i$ on site $i$ swapping places with the spin background $c_{j p \sigma} c_{j n \sigma'}$ on site $j$. This Hamiltonian is in the electron Fock state representation with the background determined by the bilayer Heisenberg model (\ref{HJ}). Historically the spin singlet-triplet basis turned out to be convenient in treating the bilayer Heisenberg model, and consequently we will apply this representation also to the hopping term (\ref{ExcitonHopping}).

Unlike the fermionic holes in the single layer case, the exciton is composed of a fermionic doublon and holon in the same rung, and hence is a bosonic particle. The local Hilbert space on each interlayer rung is five dimensional with basis in terms of five hard-core bosons as one interlayer exciton state $| E \rangle_i$ and four different spin states. In the single-triplet basis, which is valid for both the doped and undoped case, we can introduce the four hard core-boson as one singlet state and three triplet states:
\numparts
\begin{eqnarray}
	| 0 \; 0 \rangle_i&=&
		{1\over\sqrt2}(c^\dagger_{i1\uparrow}c^\dagger_{i2\downarrow}-c^\dagger_{i1\downarrow}c^\dagger_{i2\uparrow}) |0\rangle \\
	| 1 \; 0\rangle_i&=&
		{1\over\sqrt2}(c^\dagger_{i1\uparrow}c^\dagger_{i2\downarrow}+c^\dagger_{i1\downarrow}c^\dagger_{i2\uparrow})|0\rangle \\
	| 1 \; 1 \rangle_i&=&c^\dagger_{i1\uparrow}c^\dagger_{i2\uparrow} |0\rangle\\
	| 1 \; -1 \rangle_i&=&c^\dagger_{i1\uparrow}c^\dagger_{i2\uparrow}|0\rangle.
\end{eqnarray}
\endnumparts
Then the hopping term (\ref{ExcitonHopping}) can be reexpressed as:
\begin{equation}
	H_{t,ex} = - t \sum_{<ij>} | E_j \rangle \left(  | 0 \; 0 \rangle_i \langle 0 \; 0 |_j
			+ \sum_m |1 \; m \rangle_i \langle 1  \; m |_j
			 \right) \langle E_i |.
		\label{ExcitonHop}
\end{equation}
We can introduce the total spin operator
\begin{equation}
	\mathbf{S}_i = \mathbf{s}_{i1} + \mathbf{s}_{i2}
\end{equation}
and the spin difference operator
\begin{equation}
	\widetilde{\mathbf{S}} = \mathbf{s}_{i1} - \mathbf{s}_{i2}.
\end{equation}
Explicitly in terms of singlet and triplet rung states for $S=\frac{1}{2}$, this reads
\numparts
\begin{eqnarray}
	S^z_i & = & | 1  \; 1 \rangle \langle 1 \; 1 |
		- | 1 \; -1 \rangle \langle 1 \; -1 | \\
	S^+_i & = & \sqrt{2} \left( | 1 \; 1 \rangle \langle 1 \; 0 |
		+ | 1 \; 0 \rangle \langle 1 \; -1 | \right) \\
	\widetilde{S}^z_i & = & - | 0 \; 0 \rangle \langle 1 \; 0 |
		- | 1 \; 0 \rangle \langle 0 \; 0 | \\
	\widetilde{S}^+_i & = & \sqrt{2} \left( | 1 \; 1 \rangle \langle 0 \; 0 |
		- | 0 \; 0 \rangle \langle 1 \; -1 | \right).
\end{eqnarray}
\endnumparts
In general, we see that the operator $\mathbf{S}_i$ conserves the total onsite spin, while $\widetilde{\mathbf{S}}$ always changes the total spin number $s$ by a unit. The $z$-components of the spin operators do not change the magnetic number $m$, while the $\pm$-components of the spin operators change the magnetic number by a unit. The bilayer Heisenberg model is now written as
\begin{equation}
	H_J = \frac{J}{2} \sum_{<ij>}
		\left( \mathbf{S}_i \cdot \mathbf{S}_j
			+ \widetilde{\mathbf{S}}_i \cdot \widetilde{\mathbf{S}}_j \right)
		+ \frac{J_\perp}{4} \sum_i
		\left( \mathbf{S}^2_i - \widetilde{\mathbf{S}}^2_i \right).\label{HJ1}
\end{equation}
In conclusion, we formulated the exciton t-J model in the singlet-triplet basis which will be a starting point to solve the dynamics of the single exciton.

\subsection{Sign problem}
Notice also that the Hilbert space no longer contains fermionic degrees of freedom. The question is whether the disappearance of the fermionic structure also leads to the disappearance of the fermionic \emph{sign} structure, which causes so much difficulties in the single layer $t-J$ model\cite{Wu:2008p3149}.

The sign structure can be investigated as follows. Remember that at half-filling the fermionic signs in the standard $t-J$ model on a bipartite lattice can be removed by a Marshall sign transformation\cite{Marshall:1955p5}. Upon doping, signs reappear whenever a hole is exchanged with (for example) a down spin. Which matrix elements of the Hamiltonian become positive (and thus create a minus sign in the path integral loop expansion) depends on the specific basis and on the specific Marshall sign transformation.

For the double layer exciton model, define a spin basis state with a built-in Marshall sign transformation of the form (compare to Ref. \cite{Weng:2007p3423})
\begin{equation}
	| \phi \rangle = (-1)^{N_{An}^\downarrow + N_{Bp}^\downarrow}
		\left| \cdots
		\begin{array}{ccc} \downarrow & \uparrow \downarrow & \uparrow \\
			\downarrow & 0& \downarrow \end{array}
		\cdots \right>
\end{equation}
where $N_{An}^\downarrow$ is the number of down spins on the $A$ sublattice in the $n$-layer and similary we define $N_{Bp}^\downarrow$. With these basis states the Heisenberg terms are sign-free and the only positive matrix elements come from the exchange of an exciton with a $m = \pm 1$ triplet.

We conclude that, even though the model is purely bosonic, the exciton $t-J$ model is not sign-free and it is not possible to remove this sign structure using a Marshall or similar transformation.\footnote{We are not claiming that the sign structure cannot be removed. Of course, if we would know the exact eigenstates of the Hamiltonian there would be no sign problem. However, finding a basis where the sign structure vanishes is in general a NP-hard problem \cite{Troyer2005}.} However, as will be further elaborated upon in section \ref{SectionSCBA}, for both the antiferromagnetic and singlet ground states these signs do cancel out. Therefore for such ordered bilayers the problem of exciton motion turns out to be effectively bosonic.

\section{Undoped case: the bilayer Heisenberg model}
\label{SectionHeisenberg}
Before considering the dynamics of the exciton and expressing the interaction between the exciton and the spin background, we need to derive a spin wave theory for the bilayer Heisenberg model. Similar to the traditional Holstein-Primakoff spin-wave theory, we need a classical reference state, i.e. the mean field ground state of the bilayer Heisenberg model, then develop the linear order for the spin wave theory from the mean field ground state. The method we present here is similar to the one presented in \cite{Sommer:2001p5326}.

\subsection{Mean field ground state}
The singlet-triplet basis (\ref{HJ1}) of the bilayer Heisenberg model is convenient for mean field theory. Mean field theory tells us that for large ratio $J_\perp / J$ the ground state is the singlet configuration $| 0 \; 0 \rangle$. For small $J_\perp / J$, we expect antiferromagnetic ordering, which amounts to staggered condensation of $\widetilde{S}^z$. By setting $\langle \widetilde{S}^z \rangle = (-1)^i \widetilde{m}$ we obtain a mean field Hamiltonian
\begin{equation}
	H^{MF}_J = \sum_{i} \left[ \frac{1}{4} J z \widetilde{m}^2
			+ \frac{J_\perp}{4} \left( S^2_i - \widetilde{S}^2_i \right)
			- \frac{1}{2} J z \widetilde{m} (-1)^i \widetilde{S}^z_i
		\right]
	\label{MeanField}
\end{equation}
which has a ordered-disordered transition point at
\begin{equation}
	\alpha_c \equiv
	\left( \frac{J_\perp}{Jz} \right)_{c}
		= \frac{4}{3} S (S+1)
	\label{LargeSsol}
\end{equation}
where $S$ is the magnitude of spin of the spin operator on each site. A proof of this result can be found in \ref{AppendixLargeS}.

The basic idea of a spin wave theory\cite{Anderson:1952p5128,Kubo:1952p5109,Dyson:1956p5129} is to start from this semiclassical (mean field) ground state and describe the local excitations with respect to this ground state. One can immediately infer why the Holstein-Primakoff or Schwinger approach to spin wave theories fails for the bilayer Heisenberg model. First, the mean field ground state is no longer a N\'{e}el state for finite $\alpha$. Secondly, where Holstein-Primakoff describes one and Schwinger describes two onsite spin excitations, the bilayer Heisenberg has in fact three types of excitations. This has been pointed out by Chubukov and Morr\cite{Chubukov:1995p2296}, who called the 'third' excitation the longitudinal mode.

Here we want to point out that due to the local Hilbert space and the mean field ground state as described by (\ref{MeanField}) we can 'reach' all states in the local Hilbert space with three types of excitations: a longitudinal $e^\dagger$ which keeps the magnetic number $m$ constant, and transversal $b^\dagger_{\pm}$ who change the magnetic number $m$ by either $\pm 1$. In the limit of large $S$ these excitations tend to become purely bosonic. We will take the mean field ground state of (\ref{MeanField}) and these three excitations as the starting point for the linear spin wave theory.

Finally, we must mention the obvious flaw in the above reasoning. Where we criticized earlier spin wave theories because they predicted the wrong critical value of $J_\perp/Jz$, we now apparently adopt such a 'wrong' theory since (\ref{LargeSsol}) predicts $\alpha_c = 1$ for $S = \frac{1}{2}$! Nevertheless, as we show in \ref{AppendixQCP} concerning $S=\frac{1}{2}$, the presence of spin waves changes the ground state energy which makes the disordered state more favorable even below the mean field critical $\left( \frac{J_\perp}{Jz} \right)_c$ calculated above. Hence, due to correctly taken the ground state energy shifts into account, one finds the accurate critical value for $\alpha$ consistent with numerical calculations.

\subsection{Spin wave theory}
We will now construct explicitly the spin wave theory described above for $S=\frac{1}{2}$. First, one needs to find the ground state following equation (\ref{MeanField}). In the $S=\frac{1}{2}$ case, this amounts to a competition between the singlet state $|s=0, m=0 \rangle$ and the triplet $| s=1, m=0 \rangle$. The mean field ground state on each rung is given by a linear superposition of those two,
\begin{equation}
	| G \rangle_i = \eta_i\cos \chi | 0 \; 0 \rangle_i - \sin \chi |1 \; 0 \rangle_i,
	\label{MFGroundState}
\end{equation}
which interpolates between the N\'{e}el state ($\chi=\pi/4$) and the singlet state ($\chi=0$). The onset of antiferromagnetic order can thus be viewed as the condensation of the triplet state in a singlet background.\cite{Sommer:2001p5326} With $\eta_i = (-1)^i$ alternating we have introduced a sign change between the two sublattices $A$ and $B$. The angle $\chi$ will be determined later by self-consistency conditions.

The three operators that describe excitations with respect to the ground state are
\numparts
\begin{eqnarray}
	e^\dagger_{i} & =  &\left( \eta_{i}\sin \chi | 0 \; 0 \rangle_i
			+ \cos \chi | 1 \; 0 \rangle_i \right) \langle G |_i , \label{SW1} \\
	b^\dagger_{i+} & = & | 1 \; 1 \rangle_i \langle G | _i  ,\label{SW2}\\
	b^\dagger_{i-} & =  &| 1 \; -1 \rangle_i \langle G | _i. \label{SW3}
\end{eqnarray}
\endnumparts
The $e$-operators will later turn out to represent the longitudinal spin waves, whereas the $b$-operators represent the two possible transversal spin waves.

The bilayer Heisenberg model can be rewritten in terms of these operators. For completeness we include the parameter $\lambda$ that enables a comparison with the Ising limit ($\lambda=0$) with the Heisenberg limit ($\lambda=1$),
\begin{equation}
	\mathbf{S}_1 \cdot \mathbf{S}_2 	
		= S^z_1 S^z_2 + \frac{1}{2} \lambda ( S^+_1 S^-_2 + S^-_1 S^+_2 ).
\end{equation}
Given this, we can explicitly write down the spin operators in terms of the new $e$ and $b$ operators, as is done in \ref{AppendixBHM}.

From the requirement that the Hamiltonian does not contain terms linear in spin wave operators we obtain the self-consistent mean field condition for the ground state angle $\chi$,
\begin{equation}
	 ( \cos 2 \chi - \alpha \lambda ) \sin 2 \chi = 0
\end{equation}
which has two possible solutions. Either $\chi = 0$, which corresponds to a singlet ground state configuration, the \emph{disordered phase}. If $\cos 2 \chi = \alpha \lambda$, there exists an \emph{antiferromagnetic ordered phase}. These are indeed the two phases represented in Figure \ref{FigHeisenbergPD}. Which of the two solutions ought to be chosen, depends on the ground state energy competition. In \ref{AppendixQCP} we compare the ground state energy of both phases, from which we can deduce that the critical point lies at $\alpha_c \approx 0.6$, consistent with numerical literature\cite{Sandvik:1995p5136,Sandvik:1994p3011}.

The dispersion of the spin wave excitations can be found when if we consider only the quadratic terms in the Hamiltonian. This is called the \emph{`linear'} spin wave approximation, and it amounts to neglecting the cubic and quartic interaction terms. First take a Fourier transform of the spin wave operators
\begin{equation}
	e^\dagger_{i \sigma}
		= \sqrt{ \frac{2}{N} } \sum_{k} e^\dagger_{k\sigma} e^{i k \cdot r_i}
\end{equation}
where the sum over $k$ runs over the $2/N$ momentum points in the domain $[-\pi,\pi] \times [-\pi,\pi]$ and $\sigma = A,B$ represents the sublattice index. A similar definition is used for the $b$-operators.

Upon Fourier transformation, we can decouple the spin waves from the two sublattices $A$ and $B$ by introducing
\begin{equation}
	e^\dagger_{k,p}
		= \frac{1}{\sqrt{2}} ( e^\dagger_{kA} + p e^\dagger_{kB} )
	\label{PhaseEqn}
\end{equation}
where $p = \pm$ stand for the phase of the spin mode. Modes with $p=-1$ are out-of-phase and have the same dispersion as the in-phase $p=1$ modes but shifted over the antiferromagnetic wavevector $Q = (\pi, \pi)$. Again similar considerations hold for the $b$ operators.

Next we perform the Bogolyubov transformation on the magnetic excitations,
\numparts
\begin{eqnarray}
	\label{Btrafos1}
	e^\dagger_{k,p} & = & \cosh \varphi_{k,p} \zeta^\dagger_{k,p} + \sinh \varphi_{k,p} \zeta_{-k,p} \label{BTrafo1} \\
	b^\dagger_{k,p,+} & = & \cosh \theta_{k,p} \alpha^\dagger_{k,p} + \sinh \theta_{k,p} \beta_{-k,p} \\
	b^\dagger_{k,p,-} & = & \cosh \theta_{k,p} \beta^\dagger_{k,p} + \sinh \theta_{k,p} \alpha_{-k,p}
	\label{Btrafos3}
\end{eqnarray}
\endnumparts
The corresponding transformation angles are set by the requirement that the Hamiltonian becomes diagonal in the new operators $\zeta$ (the longitudinal spin wave) and $\alpha, \beta$ (the transversal spin wave). In doing so, we introduced the \emph{'ideal'} spin wave approximation in which we assume that the spin wave operators obey bosonic commutation relations\cite{Dyson:1956p5129}. This assumption is exact in the large $S$ limit. For $S=\frac{1}{2}$ this approximation turns out to work extremely well\cite{Manousakis:1991p2291}, since the corrections to the bosonic commutation relations are expressed as higher order spin-wave interactions. The Bogolyubov angles are therefore given by
\begin{eqnarray}
	\tanh 2\varphi_{k,p} &= &
		\frac{ -p \frac{1}{2} \cos^2 2 \chi \gamma_k}
			{\sin^2 2\chi + \lambda \alpha \cos 2\chi
			-p {1\over2} \cos^2 2 \chi \gamma_k} ,
	\label{Bogo1}
	 \\
	\tanh 2\theta_{k,p} &=&
		\frac{p \lambda \gamma_k}
			{\sin^2 2\chi+ (1+\lambda) \alpha \cos^2 \chi
			-p \lambda \cos 2\chi\gamma_k}.
	\label{Bogo2}
\end{eqnarray}
The factor $\gamma_k$ encodes for the lattice structure, and it equals for a square lattice
\begin{equation}
	\gamma_k = \frac{1}{z} \sum_{\delta} e^{i k \cdot \delta}
		= \frac{1}{2} \left( \cos k_x + \cos k_y \right)
\end{equation}
where the sum runs over all nearest neighbor lattice sites $\delta$.
The Bogolyuobov angles still depend on $\chi$, which characterizes the ground state. In the antiferromagnetic phase $ \cos 2 \chi = \lambda \alpha$ and for the Heisenberg limit $\lambda=1$ these angles reduce to
\begin{eqnarray}
	\tanh 2\varphi_{k,p} &= &
		\frac{-p \alpha^2 \gamma_k}{ 2 - p \alpha^2  \gamma_k } ,
	 \\
	\tanh 2\theta_{k,p} &=&
		\frac{p  \gamma_k}{ 1+\alpha - p \alpha  \gamma_k }.
\end{eqnarray}
We can distinguish between the longitudinal and transversal spin excitations, with their dispersions given by
\begin{eqnarray}
	\epsilon^L_{k,p} & = &
	Jz \sqrt{1 - p \alpha^2 \gamma_k } \\
	\epsilon^T_{k,p} & = &
	\frac{1}{2} Jz \sqrt{ ( 1 + \alpha ( 1 - p \gamma_k) )^2 - \gamma_k^2 }
\end{eqnarray}
The longitudinal spin wave is gapped and in the limit where the layers are decoupled ($\alpha=0$) completely non-dispersive, while the transversal spin wave is always linear for small momentum $k$. This type of spectrum is similar to a phonon spectrum, which contains a linear $k$-dependent acoustic mode and a gapped flat optical mode. This correspondence between spin waves and phonons enables us to use techniques from electron-phonon interaction studies for the exciton-spin wave interactions.

On the other hand, in the singlet phase ($\alpha > 1$) one has trivially three identical triplet spin excitations. The Bogolyubov angles are given by
\begin{equation}
	\tanh 2\varphi_{k,p}
	= - \tanh 2\theta_{k,p}
	= \frac{- p \gamma_k}{ 2 \alpha - p \gamma_k }
\end{equation}
and the dispersion of the triplet spin waves is
\begin{equation}
	\epsilon_{k,p} = Jz \sqrt{\alpha ( \alpha - p \gamma_k) }.
\end{equation}
These dispersions correspond to earlier numerical and series expansions results\cite{Kotov:1998p5125,Weihong:1997p5123,Gelfand:1996p5121,Chubukov:1995p2296}. In fact, these results are exactly equal to the dispersions obtained from the non-linear sigma model\cite{Duin:1997p2301}.

The above derivation adds to earlier studies of the bilayer Heisenberg model in that we now found explicit expressions of how the spin waves are related to local spin flips, equations (\ref{BTrafo1})-(\ref{Bogo2}). This microscopic understanding of the magnetic excitations of the system enables us in the next section to derive exactly how magnetic interactions influence the dynamics of excitons.

\begin{figure}
 \includegraphics[width=12cm]{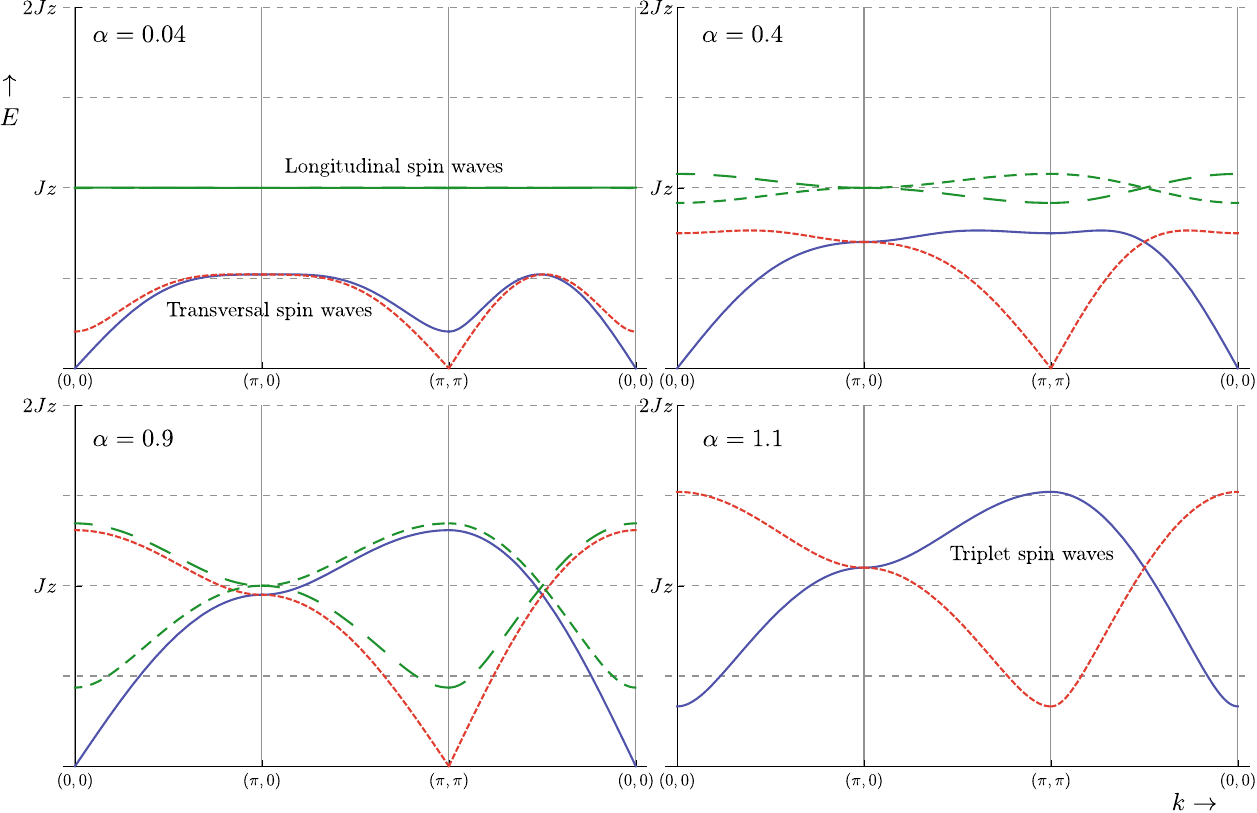}
 \caption{\label{FigModes}Dispersion of the bilayer Heisenberg spin waves for different values of $\alpha$. The top row has $\alpha=0.04$ and $\alpha=0.4$, the bottom row $\alpha=0.9$ and $\alpha=1.1$. In the antiferromagnetic phase (first three pictures) there is a clear distinction between the longitudinal spin waves (long dashed lines in green) and the transversal spin waves (solid line in blue; and the short dashed in red). The first is gapped, whilst the latter is zero at either $k = (0,0)$ or $(\pi,\pi)$ with a linear energy-momentum dependence. In the singlet phase, all spin waves are gapped triplet excitations (depicted as solid blue line and dashed red line).}
\end{figure}

\section{A single exciton in a correlated bilayer}
\label{SectionSCBA}
As was pointed out in section \ref{ExcitonParagraph}, the exciton $t-J$ model is still troubled by the sign problem even though it is purely bosonic. The sign-problem makes it difficult to say anything conclusive for systems with a finite density of excitons. Doping the single layer $t-J$ model leads to similar loss of theoretical control, and is the consequence of the fact that the magnetic ground state changes rapidly with doping. However, we can derive the dynamics of a \emph{single} exciton in the undoped bilayer. In the thermodynamic limit a single exciton will not change the ground state. Following the exciton hopping Hamiltonian (\ref{ExcitonHopping}) we can express the dynamics of the exciton upon interaction with the spin wave modes. A single exciton can be physically realized by either exciting a interlayer charge-transfer exciton in the undoped bilayer, or by infinitesimal small chemical doping of layered structures.

From a theoretical side, the spin wave we derived in the last section \ref{SectionHeisenberg} can be used to constructed the effective theory of the single exciton and apply the self-consistent Born approximation. Similar to the single layer case\cite{SchmittRink:1988p10}, we consider the mean field state $|G\rangle$ as the vacuum state and it is straightforward to derive the effective theory for single exciton as:
\begin{equation}
  H_{t,ex} = t\sum_{\langle ij \rangle}E_j^\dagger E_i \left[\cos2\chi(1-e_i^\dagger e_j)
  +\sin2\chi(e_i^\dagger+e_j)
  -\sum_\sigma b_{i\sigma}^\dagger b_{j\sigma} \right] + h.c..
  \label{exciton1}
\end{equation}

The dynamics of a single exciton are contained in the dressed Greens function, formally written as
\begin{equation}
	G^p (k, \omega) = \langle \psi_0 | E_{k,p} \frac{1}{\omega - H + i \epsilon} E_{k,p}^\dagger | \psi_0 \rangle	
\end{equation}
where $E^\dagger_{k,p}$ is the Fourier transformed exciton creation operator, and $p$ indicates the same phase index as used for the spin waves in equation (\ref{PhaseEqn}). The $|\psi_0\rangle$ denotes the ground state that arises from the spin wave approximation\cite{Manousakis:1991p2291}, that is: it is defined by the conditions
\begin{equation}
	\zeta_{k,p} | \psi_0 \rangle 
	= \alpha_{k,p} | \psi_0 \rangle
	= \beta_{k,p} | \psi_0 \rangle
	= 0
\end{equation}
for all $k,p$. Note that $|\psi_0 \rangle$ is not equal to the mean field ground state $| G \rangle$ defined in equation (\ref{MFGroundState}).

Now the Greens function cannot be solved exactly and one needs to write out a diagrammatic expansion in the parameter $t$. For this purpose, we have derived the corresponding Feynman rules of the exciton $t-J$ model in \ref{AppendixHopping}.

Using Dyson's equation one can rephrase the diagrammatic expansion in terms of the self-energy $\Sigma^p (k, \omega)$ such that
\begin{equation}
	G^p(k, \omega) = \frac{1}{\omega - \epsilon_0^p (k) - \Sigma^p (k, \omega) + i \epsilon}
\end{equation}
where $\epsilon_0^p (k)$ is the dispersion in the absence of spin excitations for the exciton with phase $p$. The self-energy can be computed by summing all one-particle irreducible Feynman diagrams. The degree to which exciton motion contains a free part grows with $\alpha$, and indeed the free dispersion is
\begin{equation}
	\epsilon_{0}^p (k) = p \; zt \cos 2 \chi \; \gamma_k
\end{equation}
where $\cos 2 \chi$ equals $\alpha \lambda$ in the antiferromagnetic phase and equals $1$ in the singlet phase.

As we noted before, the spin wave spectrum resembles a phonon spectrum. Hence we can compute the exciton self-energy using the Self-Consistent Born Approximation (SCBA)\cite{SchmittRink:1988p10,Kane:1989p585}, an approximation scheme developed for electron-phonon interactions but subsequently successfully applied to the single layer $t-J$ model. 

\begin{figure}
 \includegraphics[width=12cm]{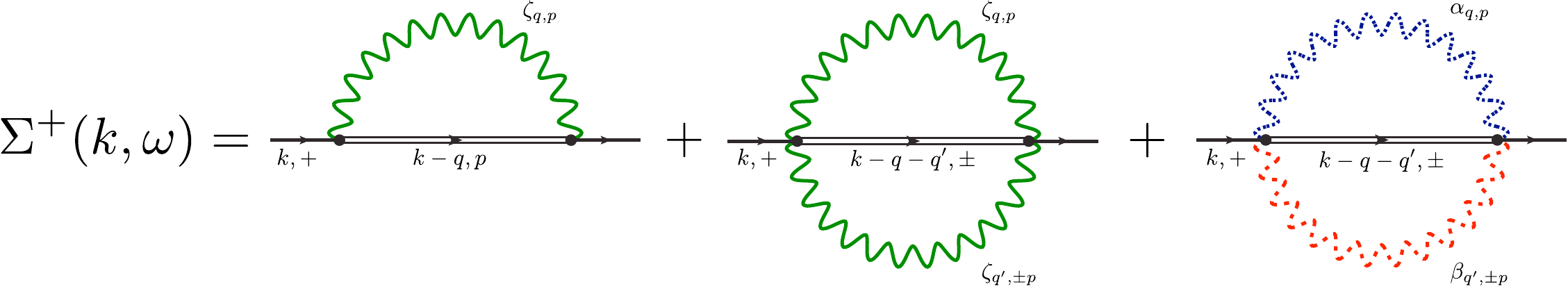}
 \caption{\label{FigSCBA}Feynman diagram representation of the Self-Consistent Born Approximation (SCBA) of equation (\ref{SCBA}). The self-energy of the exciton depends self-consistently on 'rainbow' diagrams where it emits and absorbs either one or two spin waves. The left two diagrams contain interaction with the longitudinal spin wave (solid green wavy propagators with $\zeta$ labels). The diagram to the right contains the interaction with the transversal spin waves; where the dotted (blue, upper, wavy) propagator denotes the $\alpha$ spin wave and the dashed (red, lower, wavy) propagator denotes the $\beta$ spin wave. The definitions of $\zeta, \alpha$ and $\beta$ are given in equations (\ref{Btrafos1})-(\ref{Btrafos3}). Note that vertex corrections are neglected in the SCBA.}
\end{figure}

The SCBA is based on two assumptions: 1) that one can neglect vertex corrections and 2) one uses only the bare spin wave propagators. The first assumption is motivated by an extension of Migdal's theorem\footnote{For electron-phonon interaction, higher order vertex corrections are of order $\frac{m}{M}$ where $m$ is the electron mass and $M$ is the ion mass. This justifies that for electron-phonon interactions the SCBA is right \cite{FetterWalecka}. Comparisons between the SCBA and exact diagonalization methods for the single layer $t-J$ model have shown that it is justified to neglect the vertex correction there as well \cite{MartinezHorsch1991}.}, the second by the linear spin wave approximation. Consequently, all remaining diagrams are of the 'rainbow' type which can be summed over using a self-consistent equation. The assumption that the vertex corrections are irrelevant allows us to completely resum Feynman diagrams up to all orders in $t$. The SCBA is therefore not a perturbation series expansion and consequently $t$ does not necessarily has to be a small parameter.

\begin{figure}
 \includegraphics[width=8.6cm]{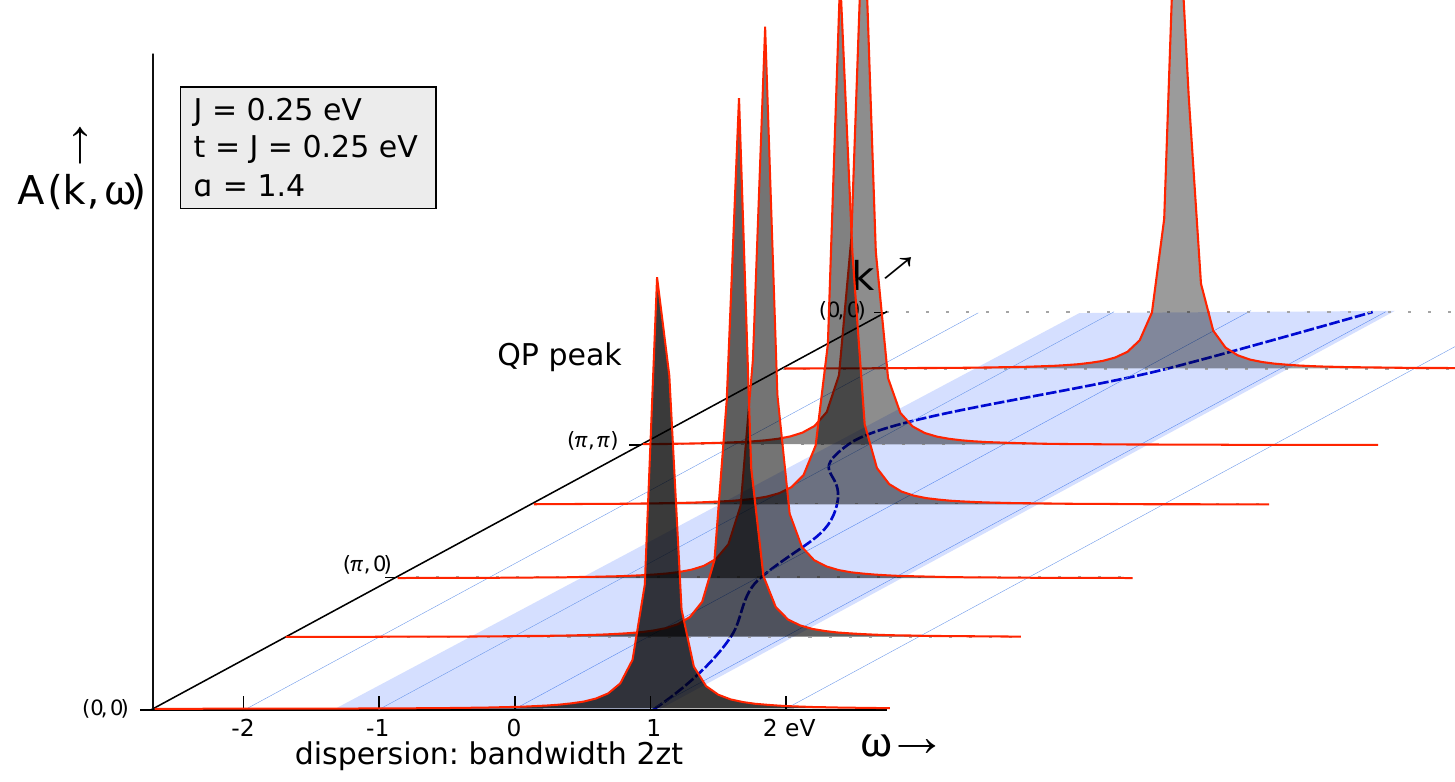}
 \caption{\label{Figalphalarge}Exciton spectral function for parameters $J=t$ and $\alpha=1.4$. The only relevant feature is the strong quasiparticle peak with dispersion equal to $8t$, where $t$ is the hopping energy of the exciton. The horizontal axis describes energy, the vertical axis is the spectral function in arbitrary units.}
\end{figure}

\begin{figure}
 \includegraphics[width=8.6cm]{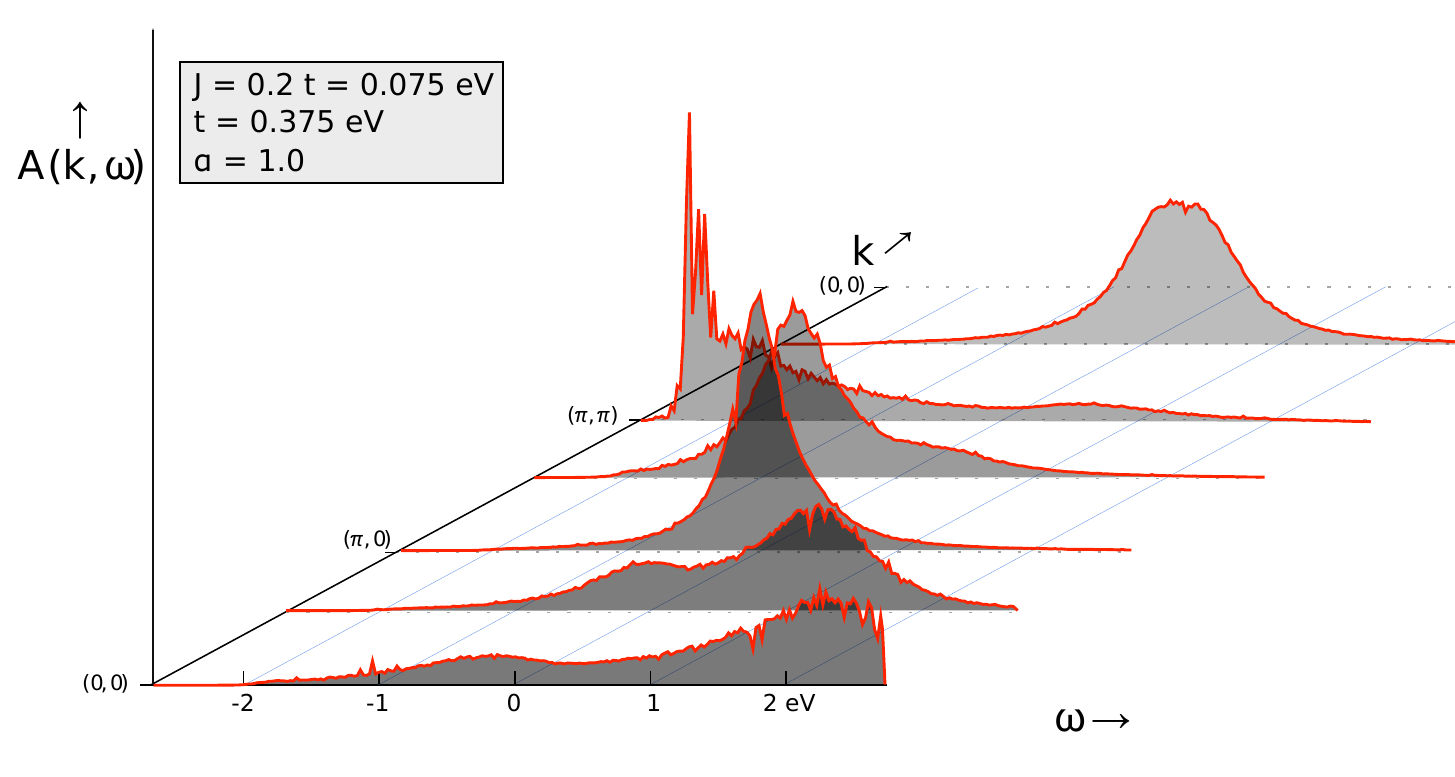}
 \caption{\label{alphaQCP}Exciton spectral function at the quantum critical point, for $J = 0.2t$ and $\alpha=1$. No distinct quasiparticle peak is observable, and at all momenta a broad critical bump appears in the spectrum.}
\end{figure}

\begin{figure}
 \includegraphics[width=12cm]{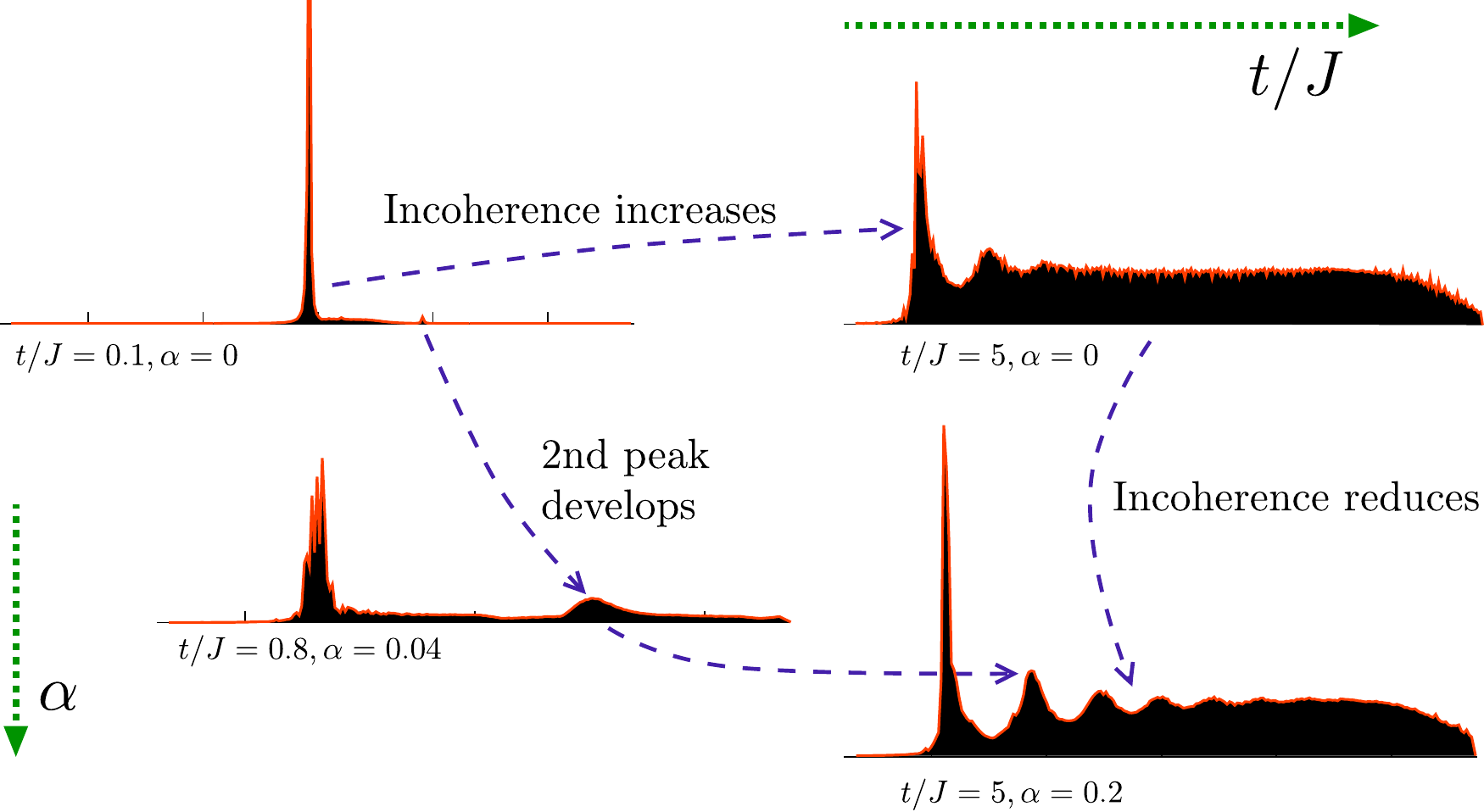}
 \caption{\label{FigOverview}A qualitative overview of zero momentum exciton spectral functions $A(k=0,\omega)$ for various parameters of $t/J$ and small interlayer coupling $\alpha$. For $\alpha$ identically zero, the ratio $t/J$ determines the amount of excited spin waves. In the adiabatic limit $t \ll J$ no spin waves can be excited by and the exciton is localized with a clear quasiparticle peak. Upon increase of $t/J$ more and more spectral weight is transferred to higher order spin wave peaks, which in the anti-adiabatic limit $t \gg J$ leads to the formation of a broad incoherent spectrum. The inclusion of a small nonzero interlayer coupling $\alpha$ reduces the incoherence of this spectrum, see equation (\ref{Ladderpeaks}). As a result the Ising-like ladder spectrum becomes more pronounced. Here we only show the zero momentum spectra, in our earlier work \cite{Rademaker2011EPL} the momentum dependence of these spectra was shown.}
\end{figure}

For the exciton $t-J$ model, the SCBA amounts to computing the self-energy for the in-phase exciton, as shown diagrammatically in Figure \ref{FigSCBA}. Usual Feynman rules dictate that we need to integrate over all intermediate frequencies of the virtual spin waves. However, under the linear spin wave approximation the spin wave propagator is $i / ( \omega' - \epsilon(k) + i \epsilon)$ which amounts to a Dirac delta function in the frequency domain integration\cite{SchmittRink:1988p10}. For example, the first diagram of Figure \ref{FigSCBA} is reduced as follows,
\begin{eqnarray}
	\frac{1}{N} \sum_{q,p} \int^\infty_{-\infty} \frac{d \omega'}{\pi}
		M_{k,q}^2 G^p (k-q, \omega-\omega')
		\left[ \frac{i}{\omega' - \epsilon^L_{k,p} + i \epsilon} \right]
	\nonumber \\
	= \frac{1}{N} \sum_{q,p}
		M_{k,q}^2 G^p (k-q, \omega-\epsilon^L_{q,p}),
\end{eqnarray}
where $M_{k,q}$ is the vertex contribution and $G^p(k,\omega)$ is the exciton propagator. Emission (or absorption) of a spin wave by an exciton can thus be incorporated by changing the momentum and energy of the exciton propagator. Analytically we write for the in-phase exciton self-energy,

\begin{eqnarray}
	\fl \Sigma^+ (k, \omega)
	 = \frac{z^2 t^2}{N} \sin^2 2\chi \sum_{q,p}
		\left( \gamma_{k-q} \cosh \varphi_{q,p} + p \gamma_k \sinh \varphi_{q,p} \right)^2
		G^p (k-q, \omega-\epsilon^L_{q,p} )
	\nonumber \\
		+ \frac{z^2 t^2}{N^2} \cos^2 2 \chi \sum_{q,q'} \sum_{\pm, p}
		( \gamma_{k+q'} \cosh \varphi_{q,p} \sinh \varphi_{q',\pm p}
	\nonumber \\
		\hspace{1cm} \pm \gamma_{k+q} \cosh \varphi_{q',\pm p} \sinh \varphi_{q,p} )^2
		G^\pm (k-q-q', \omega - \epsilon^L_{q,p} -\epsilon^L_{q',\pm p} )
	\nonumber \\
		+ \frac{z^2 t^2}{N^2} \sum_{q,q'} \sum_{\pm, p}
		( \gamma_{k-q} \cosh \theta_{q,p} \sinh \theta_{q', \pm p} 
	\nonumber \\
		\hspace{1cm} \pm \gamma_{k-q'} \cosh \theta_{q',\pm p} \sinh \theta_{q, p} )^2
		G^\pm (k-q-q', \omega - \epsilon^T_{q,p} -\epsilon^T_{q',\pm p} )
	\label{SCBA}
\end{eqnarray}

which depends on the exciton propagator and the Bogolyubov angles derived in the previous section. A similar formula to (\ref{SCBA}) applies to $\Sigma^-$. However, it is easily verified that
\begin{equation}
	\Sigma^- (k, \omega) = \Sigma^+ (k+(\pi,\pi), \omega)
\end{equation}
since $\gamma_{k+(\pi,\pi)} = -\gamma_k$. In general the SCBA (\ref{SCBA}) cannot be solved analytically, and hence we have obtained the exciton spectral function
\begin{equation}
	A (k, \omega) = - \frac{1}{\pi} \mathrm{Im} \left[ G (k, \omega) \right]
\end{equation}
using an iterative procedure with Monte Carlo integration over the spin wave momenta discretized on a 32 $\times$ 32 momentum grid. We start with $\Sigma=0$ and after approximately 20 iterations the spectral function converged. The results for typical values of $\alpha,J $ and $t$ are shown in Figures \ref{Figalphalarge} to \ref{FigYBCO}.

We start from the situation with $\alpha > 1$ where the magnetic background is a disorder phase with all spin singlet configuration in the same rung. In this case, the free dispersion of the exciton with bandwidth proportional to $t$ survived because all the magnetic triplet excitations are gapped, with an energy of $Jz \sqrt{\alpha (\alpha-1)}$. For $t<J$, the exciton-magnetic interactions will barely change the free dispersion while for $t > J$ such exciton-magnetic interactions can still occur, leading to a small 'spin polaron' effect where the exciton quasiparticle (QP) peak is diminished and spectral weight is transferred to a polaronic bump at a higher energy than the quasiparticle peak. For most values of $t/J$ this effect is however negligible already for $\alpha$ just above the critical point. The exciton spectral function for $t=J$ and $\alpha=1.4$ can be seen in Figure \ref{Figalphalarge}.

As $\alpha$ decreases towards the quantum critical point at $\alpha=1$, the gap of the triplet excitations also decreases. The effect of the exciton-magnetic interactions become more significant, which leads to an increasing transfer of spectral weight from the free coherent peak to the incoherent parts. When $\alpha$ hits the quantum critical point the gap of all spin excitations vanishes. There the motion of the exciton is strongly scattered by the spin excitations which completely destroy the coherent peak and leads to an incoherent critical hump in the spectrum as shown in Figure \ref{alphaQCP}. For $\alpha$ further decreases to values $\alpha <1$, magnetic background becomes antiferromagnetically ordered with two gapless transverse modes and one gapped longitudinal mode. In this case, the motion of the exciton is still strongly scattered with the spin excitations leaving a footprints in the exciton spectrum.

Most striking thing happens then at $\alpha=0$, when the two layers are effectively decoupled and we would expect similar behavior for an interlayer exciton as for a hole or electron in a single layer. Indeed conform with the single hole in the $t-J$ model\cite{SchmittRink:1988p10,Kane:1989p585} we find that a moving exciton causes spin frustration with an energy proportional to $J$. In the limit where $J \gg t$ the kinetic energy of the exciton is too small to be able to move through magnetic background. Therefore, we expect a localization of the exciton which is reflected in spectral data by an almost non-dispersive quasiparticle peak. This peak has a bandwidth proportional to $t^2/J$ and carries most of the spectral weight, $1 - \mathcal{O} (t^2/J^2) $. The remaining spectral weight is carried by a second peak, at an energy $Jz$ above the main peak. 

More complex behavior at $\alpha=0$ arises in the anti-adiabatic limit $t \gg J$, where the kinetic energy of the exciton is large compared to the energy required to excite (and absorb) spin waves. Consequently, many spin waves are excited as the exciton moves and the exciton becomes 'overdressed' with multiple spin waves. At nonzero $J$ however, a very small quasiparticle peak remains with a bandwidth of order $J$. Nonetheless the majority of spectral weight is carried in the incoherent many-spin wave part.

However, realistic physical systems are expected to have a small nonzero value of $\alpha$ and an intermediate value of $t/J$. What happens here? A simple extrapolation of the two aforementioned cases yields that the bandwidth of the quasiparticle peak will reach its maximum value at $J \approx t$. Similar extrapolations suggest that about half of the spectral weight will be carried by the QP peak. However, inclusion of a finite value of $\alpha$ is not so trivial on an analytical level. Numerical results are therefore needed, and an overview of spectral functions for different ratios of $t/J$ and small values of $\alpha$ is given in Figure \ref{FigOverview}.

\subsection{Development of Ising-like confinement}
Upon the inclusion of a small nonzero interlayer coupling $\alpha$ a ladder spectrum seems to appear, reminiscent of the spectrum of a single hole in a Ising antiferromagnet. Physically, this can be understood as follows. In the $\alpha=0$ limit, the magnetic interactions are dominated by the transverse excitations which are just single layer spin waves. For any finite $\alpha > 0$ the (interlayer) longitudinal spin waves become increasingly relevant. To understand their effect on the exciton spectral function, consider the SCBA equation (\ref{SCBA}), neglect the diagrams involving transversal spin waves and expand the self-energy up to first order in $\alpha$. Only the single spin wave diagram contributes and it equals
\begin{equation}
	\Sigma^+ (k, \omega) = \frac{z^2t^2}{N} \sum_{q,\pm}
		\gamma_{k-q}^2 G^{\pm} (k-q, \omega-Jz)
\end{equation}
from which we deduce, observing that $\Sigma^- = \Sigma^+$ and shifting the momentum summation, that the self-energy must be momentum-independent and given by the self-consistent equation
\begin{equation}
	\Sigma (\omega) = \frac{\frac{1}{2} z^2 t^2}{\omega - Jz - \Sigma (\omega - Jz)}.
	\label{Ladderpeaks}
\end{equation}
This self-energy is exactly the same as the self-energy of a single dopant moving through an Ising antiferromagnet\cite{Kane:1989p585}. In fact, any system where a moving particle automatically excites a gapped and flat mode the self-consistent equation (\ref{Ladderpeaks}) applies.

As described in \cite{Kane:1989p585}, a hole in an Ising antiferromagnet is effectively confined by the surrounding magnetic texture. Each hop away from its initial point increases the energy, thus creating a linear potential well for the hole. In such a linear confinement potential a ladder spectrum appears where the energy distance between the to lowest peaks scales as $t (J/t)^{2/3}$. The spectral weight carried by higher order peaks vanishes as $t/J \rightarrow 0$ \cite{Kane:1989p585}.
 
The Ising-like features in the exciton spectral function are explicitly visible in the numerically computed dispersions shown in Figure \ref{FigOverview} and Figure 2 of ref. \cite{Rademaker2011EPL}. We indeed conclude that the visibility of the ladder spectrum is actually enhanced in the bilayer case presented here relative to the hole in the single layer due to the nondispersive interlayer spin excitations. 

Of course the exciton ladder spectrum in Figure \ref{FigOverview} is not exactly sharp. By the above analysis, we can infer that the incoherent broadening of peaks is due to interactions with the transversal spin waves. Indeed, the transversal spin waves can be viewed as the equivalent of the single layer spin waves. Therefore for small $\alpha$ the effect of transversal spin waves is to reproduce the results for a single hole in the $t-J$ model, which is quasiparticle peak broadening.

\section{Relation to experiment}
\label{SectionExperiment}

The formation of bound exciton states can be experimentally verified in indirect measurements of the dielectric function or any other charge-excitation measurements. One particular example of the former is electron energy loss spectroscopy (EELS) which showed earlier clear signatures of the in-plane charge transfer excitons in cuprates\cite{Wang:1996p5177,Zhang:1998p5176}. The EELS cross-section is directly related to the dielectric function\cite{Schnatterly1979} via the dynamic structure factor $S(q, \omega)$,
\begin{equation}
	d\sigma \propto \frac{1}{q^4} S(q, \omega)
		\propto \frac{1}{q^2} \mathrm{Im}
		\left[ \frac{-1}{\epsilon(q, \omega)} \right]
\end{equation}
where the dynamic structure factor equals
\begin{eqnarray}
	S(q, \omega) &=& \frac{1}{N} \int \frac{dt}{2\pi} e^{- \epsilon |t|}
		\sum_\lambda
		\langle \psi_0 | \sum_i e^{-i q \cdot r_i}
		e^{ i (\omega - H)t}  | \lambda \rangle
	\nonumber \\ &&
		\times \langle \lambda | \sum_j e^{i q \cdot r_j} | \psi_0 \rangle 	
\end{eqnarray}
where the sum $\lambda$ runs over all intermediate states, $| \psi_0 \rangle$ is the . We use the dipole expansion such that
\begin{equation}
	e^{i q r_i}
	= 1 + i \vec{q} \cdot \vec{r}_i + \ldots	
\end{equation}
where the electron position operator can be expanded in terms of the possible electron wave functions in the tight binding approximation,
\begin{equation}
	\sum_i \vec{r}_i = \sum_{ij \sigma} c^\dagger_{i \sigma} c_{j \sigma}
		\langle \phi_{i} | \vec{r} | \phi_{j} \rangle
\end{equation}
where $| \phi_{i } \rangle$ are the Wannier wave functions of the electron on site $i$. The $z$ component of $\langle \phi_{i} | \vec{r} | \phi_{j} \rangle$ is proportional to the interlayer hopping energy $t_\perp$, which in turn is equal to the the creation operator of an exciton,
\begin{eqnarray}
	r^z & \propto & t_\perp \sum_{i \sigma} c^\dagger_{i n \sigma} c_{i p \sigma}
	 + h.c. \\
	 &\propto & t_\perp \sum_{i} \left( E^\dagger_i + E_i \right)
\end{eqnarray}
We recognize the Fourier transform of the $k=0$ excitonic state, so that we find
\begin{eqnarray}
	S(q^z, \omega) &\propto& (q^z t_\perp)^2
	\int \frac{dt}{2\pi} e^{- \epsilon |t|}
		\sum_\lambda
		\langle \psi_0 | E_{k=0} \;
		e^{ i (\omega - H)t}  | \lambda \rangle
	\nonumber \\ &&
		\times \langle \lambda | \; E_{k=0}^\dagger | \psi_0 \rangle.
\end{eqnarray}
We have introduced the term $e^{- \epsilon |t|}$ to ensure convergence of the integral so that we can integrate over $t$. We find that the dynamic structure factor is directly related to the exciton spectral function
\begin{eqnarray}
	S(q^z, \omega) &\propto& (q^z t_\perp)^2
		\langle \psi_0 | E_{k=0} \;
	\left( \frac{i}{\omega - H + i \epsilon} - \right.
	\nonumber \\ && \left.
	\frac{i}{\omega - H - i \epsilon} \right)
	\; E_{k=0}^\dagger | \psi_0 \rangle
	\nonumber \\
	& \propto & (q^z t_\perp)^2 A(k=0, \omega)
\end{eqnarray}
or in other words
\begin{equation}
	\textrm{Im} \left[ \epsilon^{-1} (q^z, \omega) \right]
	\sim (t_\perp)^2 A(k=0, \omega).
\end{equation}
Consequently, one expects that the bound exciton states to show up in EELS measurements when probing the $z$-axis excitations. In addition to the bound exciton states, a broad electron-hole continuum will show up at high energies.

Another possible way to detect interlayer excitons is to use optical probes. The optical conductivity $\sigma( q, \omega)$ of a material is related to the dielectric function\cite{BruusFlensberg} by
\begin{equation}
	\epsilon^{-1} (q, \omega) = 1 - i \frac{q^2}{\omega} V_c (q)
		\sigma( q, \omega),
\end{equation}
where $V_c(q)$ is the Fourier transform of the Coulomb potential $\frac{1}{\epsilon_0 |r-r'|}$. The real part of the $c$-axis optical conductivity is therefore proportional to the exciton spectral function. Similar considerations hold when one measures the Resonant Inelastic X-ray Scattering (RIXS)\cite{Ament:2010p5208} spectrum.

\begin{figure}
 \includegraphics[width=8.6cm]{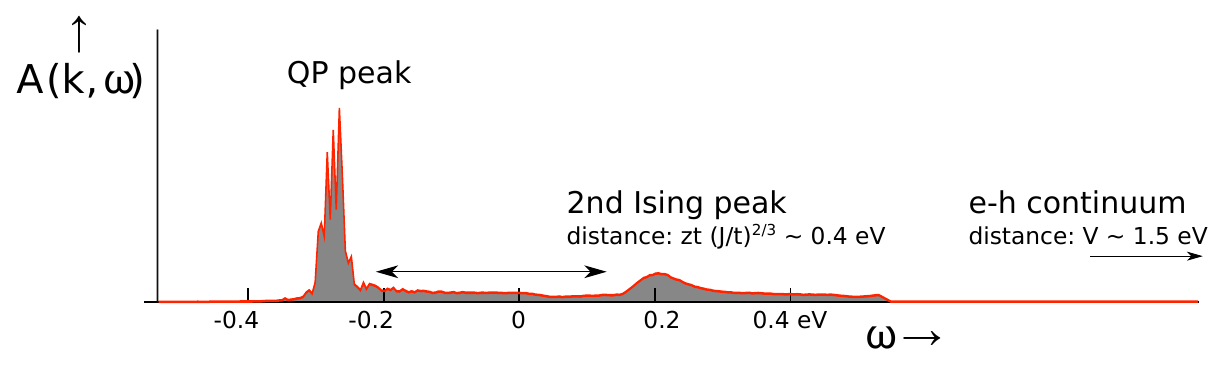}
 \caption{\label{FigYBCO}Expected zero-momentum exciton spectral function for the $c$-axis charge-transfer exciton in YBCO bilayers. We used model parameters $J=0.125$ eV, $t=0.1$ eV and $\alpha=0.04$. A pronounced quasiparticle peak is followed at a distance of $zt (J/t)^{2/3}$ by a secondary peak as a sign of Ising confinement. The electron-hole continuum sets in at an energy $V \sim 1.5$ eV above the center of this spectrum. The momentum dependence of this spectrum is shown in Ref. \cite{Rademaker2011EPL}.}
\end{figure}

When comparing the dielectric function with the computed spectral functions in Figures \ref{Figalphalarge}-\ref{FigYBCO}, bear in mind that the latter are shifted over the energy $E_0$ required to excite an interlayer exciton. This energy is of the order of electronvolts. For example, along the $ab$-plane in cuprates charge-transfer excitons are observed in the range of 1-2 eV\cite{Basov:2005p1689}. Since the energy required for a charge-transfer excitation is largely dependent on the onsite repulsion, we expect that the $c$-axis exciton will be visible at comparable energy scales.

How would then the exciton spectrum look like for a realistic material, such as the bilayer cuprate YBa$_2$Cu$_3$O$_{7-\delta}$ (YBCO)? Following earlier neutron scattering experiments\cite{Imada:1998p2790,Tranquada:1989p5209} one can deduce that the effective $J = 125 \pm 5$ meV and $J_\perp = 11 \pm 2$ meV, which corresponds to an effective value of $\alpha = 0.04 \alpha_c$ where $\alpha_c$ is the critical value of $\alpha$. \cite{Chubukov:1995p2296}. The question remains what a realistic estimate of the exciton binding energy is. The planar excitons are known to be strongly bound \cite{Zhang:1998p5176} with binding energy of the order of 1-2 eV. Since the Coulomb repulsion scales as $V\sim (\epsilon r)^{-1}$, we can relate the binding energy of the interlayer excitons to that of the planar excitons. The distance between the layers is about twice the in-plane distance between nearest neighbor copper and oxygen atoms, but simultaneously we expect the dielectric constant $\epsilon_c$ along the $c$-axis to be smaller than $\epsilon_{ab}$ due to the anisotropy in the screening. Combining these two effects, we consider it a reasonable assumption that the interlayer exciton binding energy is comparable to the in-plane binding energy. The hopping energy for electrons is approximately $t_e = 0.4$ eV which yields, together with a Coulomb repulsion estimate of $V \sim 1.5$ eV, an effective exciton hopping energy of $t \sim 0.1$ eV. Note that these estimates of $V/t$ justify our use of the strong coupling limit in section \ref{ExcitonParagraph}.

The spectral function corresponding to these parameters is shown in Figure \ref{FigYBCO}. Since $t \sim J$ the ladder spectrum is strongly suppressed compared to the aforementioned anti-adiabatic limit. However, the Ising confinement still shows its signature in a small `second ladder peak' at $0.4$ eV energy above the exciton quasiparticle peak. To the best of our knowledge and to our surprise, the $c$-axis optical conductivity of YBCO has not been measured before in the desired regime with energies above 1 eV\footnote{Confirmed in private communications with D. van der Marel. In addition, standard review articles on optical absorption in cuprates (such as \cite{Basov:2005p1689}) indeed only show infrared measurements ($< 1000 \, \mathrm{cm}^{-1}$) of the $c$-axis optical absorption in insulating cuprates.}. Detection of this second ladder peak in future experiments would suggest that indeed the interlayer excitons in cuprates are frustrated by the spin texture.

\section{Conclusion}
Using a rung linear spin wave theory for the bilayer Heisenberg model we constructed a theory of strongly bound excitons in a strongly correlated bilayer system. Surprisingly, for small but finite $\alpha = \frac{J_\perp}{Jz}$ the exciton becomes confined in a fashion similar to Ising confinement. The resulting ladder spectrum should be visible in measurements of the dielectric function, such as EELS, RIXS or optical conductivity.

Possible candidate materials are for example heterostructures of $n$ and $p$-type doped cuprates such as Nd$_{2-x}$Ce$_x$CuO$_4$/La$_{2-x}$Sr$_x$CuO$_4$. In YBCO or Bi$_2$Sr$_2$CaCu$_2$O$_{8+\delta}$, the copperoxide layers come in pairs which suggests the possibility of interlayer charge-transfer excitons. A spectrum of $c$-axis excitons in undoped YBCO is shown in Figure \ref{FigYBCO}.

Our model can be extended to different stacking structures. For example, in 214 compounds the sites in adjacent cuprate layers do not lie above each other, and we might need to include new interlayer magnetic interactions such as the Moriya-Dzyaloshinskii interaction. Different lattice structures can also be studied, of which the hexagonal lattice (as in graphene) is the most relevant.

One may wonder to what extent the used approximations are generally valid, such as the linear and ideal spin wave approximation. For the single layer Heisenberg model, it was shown that the next-to-leading order corrections where indeed significantly smaller\cite{Manousakis:1991p2291}, justifying the use of both approximations in that case. Together with the fact that we were able to reproduce the known phase diagram and excitation spectrum, this suggests our approach for the bilayer Heisenberg model is justifiable. Nevertheless, an exact computation of the next-to-leading order corrections can quantify the errors of the used spin wave approximations.

Another approximation we used was the expansion in large $V$, the exciton coupling strength. This coupling originates in the interlayer Coulomb interaction, from which we only consider the on-site and nearest neighbor terms. Therefore our model cannot describe accurately the process of how excitons are formed out of separate doublons and holons. We think this is a very interesting open question, especially at finite temperatures. In addition, the formation process is also accompanied by an exciton annihilation process which we neglected in our current work.

Besides the interesting properties of the exciton formation process, we think that further research should be directed towards finite densities of excitons\cite{Ribeiro2006}. The dynamical spin-hole frustration effects that are well known in the context of doped Mott insulators occur
in a strongly amplified form dealing with interlayer excitons in Mott-insulating bilayer systems. This gives further impetus to the pursuit to create such finite density correlated exciton systems
in the laboratory. One can wonder whether such physics is already at work in the  four-layer material Ba$_2$Ca$_3$Cu$_4$O$_8$F$_2$ where self-doping effects occur creating simultaneously $p$ and $n$-doped layers\cite{Chen:2006bs}. Much effort has been devoted to create equilibrium finite exciton densities
using conventional semiconductors\cite{Moskalenko:2000p4767}, while exciton condensation has been demonstrated in  coupled semiconductor
2DEGs \cite{Eisenstein:2004go,Butov:2007fr}. In strongly correlated heterostructures, however, formation of finite exciton densities is still far from achieved, although recent developments on oxide interfaces indicate exciting potential (see for example \cite{Pentcheva:2010p5025}). Besides the closely coupled $p$- and $n$-doped conducting interface-layers in these SrTiO$_3$-LaAlO$_3$-SrTiO$_3$ heterostructures, further candidates would be closely coupled $p$- and $n$-doped cuprates, such as YBa$_2$Cu$_3$O$_{7-x}$ or La$_{2-x}$Sr$_x$CuO$_4$ with Nd$_{2-x}$Ce$_x$CuO$_4$. The feasibility of this has already been experimentally demonstrated, e.g. in \cite{Takeuchi95}, but the exact interface effects need to be investigated in more detail, both experimentally as well as theoretically \cite{Ribeiro2006,Millis:2010p5231}.

\ack

This research was supported by the Dutch NWO foundation through a VICI grant. The authors wish to thank Hans Hilgenkamp, Jeroen van der Brink, Sergei Mukhin, Matthias Vojta and Dirk van der Marel for helpful discussions.

\appendix

\section{Large $S$ limit bilayer Heisenberg model}
\label{AppendixLargeS}
In this appendix we will prove equation (\ref{LargeSsol}). The mean field Hamiltonian (\ref{MeanField}) depends on the antiferromagnetic (AF) order parameter $\widetilde{m}$. We must find the ground state energy of (\ref{MeanField}) as a function of $\widetilde{m}$ and then minimize with respect to $\widetilde{m}$, thus yielding the mean field value of the AF order parameter.

However, since we are only interested in the critical value $\alpha_c$ where $\widetilde{m}$ changes from nonzero to zero, we can proceed as follows. In the singlet phase ($\widetilde{m}=0$) the mean field Hamiltonian is reduced to
\begin{equation}
	H^{(0)} = J_\perp S_1 \cdot S_2
\end{equation}
which has as ground state the singlet $| 0 \; 0 \rangle$ and as first excited state the triplet $| 1 \; 0 \rangle$ with energy difference $E_1 - E_0 = J_\perp$. We will treat the Hamiltonian terms that depend on $\widetilde{m}$ as a perturbation, and compute the ground state energy in second order perturbation theory for small $\widetilde{m}$. If the ground state energy decreases with nonzero $\widetilde{m}$, then there is an instability towards antiferromagnetism. The perturbation Hamiltonian is
\begin{equation}
	H^{(1)} = \frac{1}{4} Jz \widetilde{m}^2 - \frac{1}{2} Jz \widetilde{m} (-1)^i
		\widetilde{S}^z
\end{equation}
and the first and second order corrections to the ground state energy are
\begin{equation}
	E^{(1)}_0 + E^{(2)}_0
	= \langle 0 \; 0 | H^{(1)} | 0 \; 0 \rangle
	 + \sum_{s=1}^{2S}
	 	\frac{|\langle s \; 0| H^{(1)} | 0 \; 0 \rangle|^2}
		{E_0^{(0)} - E_s^{(0)}}.
\end{equation}
Now $H^{(1)}$ contains one term that is just an identity operator, and the $\widetilde{S}^z$ operator can only change the total spin number $s$ by one single unit. This means that the former expression yields
\begin{eqnarray}
	E^{(1)}_0 + E^{(2)}_0
	&=& \frac{1}{4} Jz \widetilde{m}^2
	 - \frac{(Jz)^2 \widetilde{m}^2}
		{4 J_\perp} \;
		|\langle 1 \; 0| \widetilde{S}^z | 0 \; 0 \rangle|^2
	\nonumber \\
	&=& \frac{Jz \widetilde{m}^2}{4 \alpha}  \left[
		\alpha - |\langle 1 \; 0| \widetilde{S}^z | 0 \; 0 \rangle|^2
		\right].
\end{eqnarray}
We see that whenever $\alpha > |\langle 1 \; 0| \widetilde{S}^z | 0 \; 0 \rangle|^2$, the ground state energy always increases when $\widetilde{m}$ is nonzero. Hence the critical value of $\alpha$ is given by
\begin{equation}
	\alpha_c = |\langle 1 \; 0| \widetilde{S}^z | 0 \; 0 \rangle|^2.
\end{equation}
The right hand side can be evaluated explicitly using Clebsch-Gordan coefficients, since
\begin{eqnarray}
	\langle 1 \; 0| \widetilde{S}^z | 0 \; 0 \rangle
	&=& \sum_{m=-S}^S 2 m \; C^{S S 1}_{m, -m , 0} \; C^{SS 0}_{m, -m, 0}
	\nonumber \\
	&=& \frac{2}{2S+1} \sqrt{ \frac{3}{S (S+1)}}  \sum_{m=-S}^S m^2
	\nonumber \\
	&=& \frac{2}{\sqrt{3}} \sqrt{S(S+1)}	
\end{eqnarray}
from which we indeed conclude that
\begin{equation}
	\alpha_c = \frac{4}{3} S (S+1).
\end{equation}

\section{Bilayer Heisenberg Hamiltonian in terms of $e,b$ operators}
\label{AppendixBHM}
The bilayer Heisenberg operators (total spin and spin difference) can be expressed in terms of the local spin excitations $e^\dagger$ and $b^\dagger$, by
\begin{eqnarray}
	\fl	S^z_{i \sigma} = b^\dagger_{+i\sigma} b_{+i\sigma} - b^\dagger_{-i\sigma} b_{-i\sigma} \\
	\fl S^+_{i \sigma} = \sqrt{2} \left(
		- \sin \chi ( b^\dagger_{+i\sigma} + b_{-i\sigma} )
		+ \cos \chi ( b^\dagger_{+i\sigma} e_{i\sigma} + e^\dagger_{i\sigma} b_{-i\sigma} ) \right) \\
	\fl \widetilde{S}^z_{i\sigma} = (-1)^{\sigma_i} \left( \sin 2 \chi ( 1 - \sum_\pm b^\dagger_{\pm i\sigma} b_{\pm i\sigma} - 2 e^\dagger_{i\sigma} e_{i\sigma} )
		- \cos 2 \chi ( e_{i\sigma}^\dagger + e_{i\sigma} ) \right) \\
	\fl \widetilde{S}^+_{i\sigma} = \sqrt{2} (-1)^{\sigma_i} \left(
		 \cos \chi ( b^\dagger_{+i\sigma} - b_{-i\sigma} )
		 + \sin \chi ( b^\dagger_{+i\sigma} e_{i\sigma} - e^\dagger_{i \sigma} b_{-i\sigma} ) \right)
\end{eqnarray}
where $\sigma$ represents the sign of the sublattice of site $i$. Consequently, the bilayer Heisenberg model in terms of these new operators reads (with $\alpha \equiv \frac{J_\perp }{Jz}$ and $\sigma = A,B$ denotes the sublattice index),
\begin{eqnarray}
	\fl H  =   \frac{1}{4} Jz N (- \alpha - 2 \lambda \alpha \cos 2 \chi - \sin^2 2 \chi )
		+ \frac{1}{2} Jz \sum_{i} ( \cos 2 \chi - \alpha \lambda ) \sin 2 \chi
			(e_{i \sigma}^\dagger + e_{i \sigma} )
	\nonumber \\ 
		+ Jz \sum_{i} (\sin^2 2 \chi + \alpha \lambda \cos 2 \chi )
			e^\dagger_{i\sigma} e_{i\sigma}
	\nonumber \\ 
		- \frac{1}{2} J \sum_{i \in A, \delta} \cos^2 2 \chi
			( e^\dagger_{iA} + e_{iA} )( e^\dagger_{i+\delta,B} + e_{i+\delta, B})
	\nonumber \\ 
		+ \frac{1}{2} J z \sum_{i \pm}
			( \alpha + \sin^2 2 \chi + \lambda \alpha \cos 2 \chi)
			b^\dagger_{\pm i \sigma} b_{\pm i \sigma}
	\nonumber \\ 
		+ \frac{1}{2} J \lambda \sum_{i \in A, \delta}
			\left( b_{+iA}^\dagger b_{-,i+\delta,B}^\dagger + b_{-iA} b_{+,i+\delta,B}
	\right. \nonumber \\ \left. \hspace{1cm}
			- \cos 2 \chi ( b^\dagger_{+iA} b_{+,i+\delta,B}
				+ b_{-iA} b^\dagger_{-,i+\delta,B} ) + h.c. \right)
	\nonumber \\ 
	 	+ \mathcal{O} ( b^\dagger b^\dagger e + e^\dagger b b )
		+ \mathcal{O} \left( [e^\dagger e + b^\dagger b]^2 \right)
		\label{BilayerHMagnon}	
\end{eqnarray}

We explicitly neglect the interaction terms, which are cubic and quartic in the spin wave operators. The above Hamiltonian contains a constant term (depends only on $\alpha$, $\chi$ and $\lambda$) that describes the ground state energy competition between the singlet and antiferromagnetic phase, see \ref{AppendixQCP}. The term linear in spin operators gives us the self-consistent condition for $\chi$. The quadratic terms will be diagonalized using the Fourier and Bogolyubov transformation as described in the main text.

\section{Quantum phase transition}
\label{AppendixQCP}

The ideal spin wave approximation introduces a shift in the ground state energy, similar to that in the single layer Heisenberg model\cite{Manousakis:1991p2291}. However, in the bilayer model there will be a competition between the ordered phase ($\cos 2 \chi = \alpha \lambda$) and the disordered phase ($\chi = 0$). Note for $\alpha \lambda > 1$ we automatically end up in the disordered phase.

\begin{figure}
 \includegraphics[width=6cm]{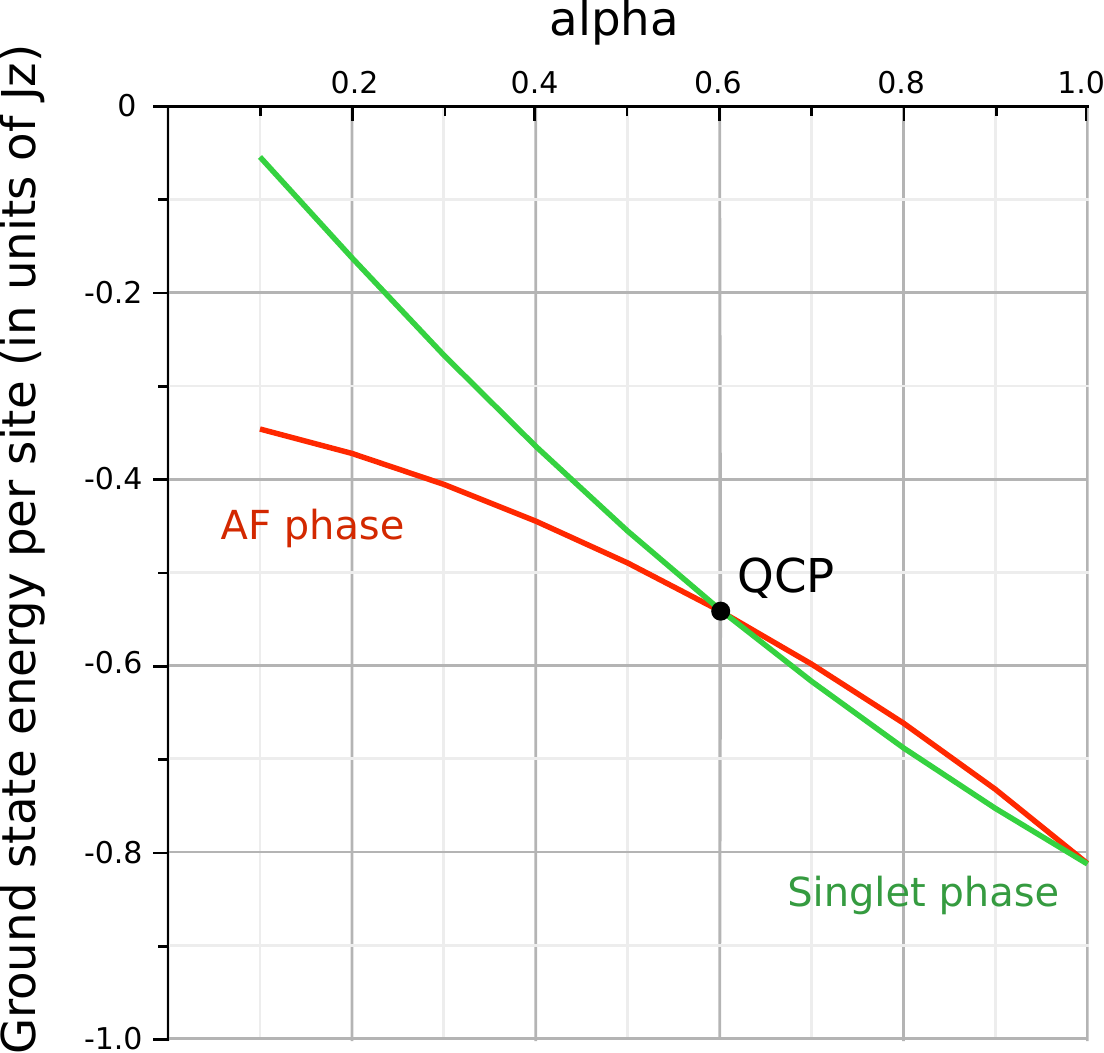}
 \caption{\label{FigGSE}Ground state energies of the bilayer Heisenberg model following equation (\ref{GroundStateEnergy}). Shown is the energy of the antiferromagnetic phase (in red) and the singlet phase (in green) for the isotropic $\lambda = 1$ model. The energies are measured in units of $JzN$. At $\alpha \approx 0.605$ there is a phase transition from the AF to the singlet phase.}
\end{figure}

For $\alpha \lambda < 1$, the ground state energy of both phases is given by the expression

\begin{eqnarray}
	\fl E_0 =
		\frac{1}{4} J z N (- \alpha  - 2 \alpha \lambda \cos 2 \chi - \sin^2 2 \chi)
	\nonumber \\ 
		+ Jz \sum_k \left[ (\sin^2 2 \chi + \alpha \lambda \cos 2 \chi) \sinh^2 \varphi_k
	\right. \nonumber \\ \left. \hspace{1cm}
			- \frac{1}{4}\cos^2 2\chi \gamma_k ( \cosh 2 \varphi_k - \sinh 2 \varphi_k ) \right]
	\nonumber \\ 
		+ Jz \sum_k \left[
			(\alpha + \sin^2 2 \chi + \alpha \lambda \cos 2 \chi ) \sinh^2 \theta_k
	\right. \nonumber \\ \left. \hspace{1cm}
			- \frac{1}{2} \lambda \gamma_k
				( \cos 2 \chi \cosh 2 \theta_k + \sinh 2 \theta_k ) \right]
		\label{GroundStateEnergy}
\end{eqnarray}

where we have to fill in the right values of $\chi$, $\theta_k$ and $\varphi_k$ depending on the phase. As can be seen in figure \ref{FigGSE}, the spin waves drive the system earlier into the singlet phase, namely at $\alpha_c \approx 0.605$. For smaller values of $\lambda$ this critical value increases, proportional to $\lambda^{-1}$.

The critical value $\alpha_c = 0.605$ for our spin wave theory closely resembles the numerical results of $\alpha_c = 0.63$. Since this ground state energy competition the system is driven into the disordered state for a different $\alpha$ than mean field theory suggests, we should replace bare values of $\alpha = \frac{J_\perp}{Jz}$ by the renormalized  $\alpha^* = \alpha / \alpha_c$ when computing the exciton spectral function.

\section{Explicit expressions for exciton-spin wave interactions}
\label{AppendixHopping}
We can rewrite the Hopping Hamiltonian (\ref{ExcitonHop}) from the singlet-triplet basis using the local spin excitation operators defined in equations (\ref{SW1})-(\ref{SW3}),

\begin{equation}
	H_t = t \sum_{\langle ij \rangle} E_j^\dagger E_i
	\left( \cos 2 \chi ( 1- e^\dagger_i e_j ) 
	+ \sin 2 \chi (e_i^\dagger + e_j)
		- \sum_\sigma b^\dagger_{i \sigma} b_{j \sigma} \right)
\end{equation}
where $\sigma$ is the sum over spins $\pm 1$ and $<ij>$ denotes nearest neighbor pairs. We need to rewrite this in terms of the longitudinal ($\zeta$) and transversal ($\alpha$ and $\beta$) modes derived in the main text. Therefore we first split all operators into the ones that live on sublattice $A$ and the ones that live on $B$,
\begin{eqnarray}
	\fl H_t = t \sum_{i \in A, \delta} E_{i+\delta,B}^\dagger E_{i,A}
		\left( \cos 2 \chi ( 1- e^\dagger_{i,A} e_{i+\delta,B} ) 
	\right. \nonumber \\ \left. \hspace{1cm}
		+ \sin 2 \chi (e_{i,A}^\dagger + e_{i+\delta,B})
		- \sum_\sigma b^\dagger_{i,A, \sigma} b_{i+\delta,B, \sigma} \right) + h.c.
\end{eqnarray}
As described in the main text, the Fourier transform for the sublattice operators is $E^\dagger_{i, A} = \sqrt{\frac{2}{N}} \sum_{k} E^\dagger_{kA} e^{ikr_i}$ and we introduce the in-phase $p=1$ and out-phase $p=-1$ exciton operators
$ E^\dagger_{k,p}
	=  \frac{1}{\sqrt{2}} ( E^\dagger_{kA} + p E^\dagger_{kB} ) $; similar expressions hold for the spinon operators. The hopping Hamiltonian now can be written as
\begin{eqnarray}
	H_0 & = &
		zt \cos 2 \chi \sum_{k,p} p \; \gamma_k \; E^\dagger_{k,p} E_{k,p} \\
	H_1 & = &
		\frac{zt}{\sqrt{N}} \sin 2 \chi \sum_{k,q} \sum_{p,p'} p
			E^\dagger_{k+q,p} E_{k,pp'} (\gamma_{k+q} e^\dagger_{-q,p'} + p' \gamma_k e_{q,p'} ) \\
	H_2^L & = &
		- \frac{zt}{N} \cos 2\chi \sum_{k,k',q} \sum_{p,p'} \sum_{\pm}
			p p' \gamma_{k-k'+q}
			E^\dagger_{k+q,p} E_{k,\pm p} e^\dagger_{k'-q,\pm p'} e_{k',p'} \\
	H_2^T & = &
		- \frac{zt}{N} \sum_{k,k',q} \sum_{\sigma} \sum_{p,p'} \sum_{\pm}
			p p' \gamma_{k-k'+q}
			E^\dagger_{k+q,p} E_{k,\pm p} b^\dagger_{k'-q,\pm p',\sigma} b_{k',p',\sigma}
\end{eqnarray}
Note that this Hamiltonian contains four different type of processes. The first line $H_0$ contains a free part of the exciton motion. The bandwidth of the free exciton dispersion increases linearly in $\alpha$ in the antiferromagnetic phase until it saturates at $2 z t$ in the disordered phase. The next term $H_1$ describes the creation and annihilation of a single longitudinal mode due to exciton motion. This term is only present in the antiferromagnetic phase and is comparable to the hole-spin vertex in the single layer $t-J$ model. Finally, there are two $H_2$ interactions where an exciton scatters off a transversal ($H_2^T$) or longitudinal ($H_2^L$) mode. These processes can also be changed into the creation or annihilation of a pair of spin modes. All processes can be characterized by a conservation of total phase index $p$ and conservation of total momentum.

The remaining step is to write out the interaction vertices explicitly in terms of the Bogolyubov transformed spin waves. The single-magnon process equals
\begin{eqnarray}
	\fl H_1 =
		\frac{zt\sin2\chi}{\sqrt N}\sum_{k_1 \ldots k_3} \sum_{p_1 \ldots p_3}
		\delta^{(2)} (k_1 - k_2 - k_3)
		\delta( \prod_{i=1}^3 p_i - 1 )
		\; E_{k_1p_1}^\dagger E_{k_2p_2}\zeta_{k_3p_3}
	\nonumber \\
		\times p_1  (p_3 \gamma_{k_2} \cosh \varphi_{k_3p_3} + \gamma_{k_1} \sinh \varphi_{k_3 p_3} )
		+h.c.
\end{eqnarray}
The process that involves two longitudinal spin waves is given by
\begin{eqnarray}
	\fl H_2^L =
		- \frac{zt \cos 2\chi}{N} \sum_{k_1 \ldots k_4} \sum_{p_1 \ldots p_4}
		\delta^{(2)} (k_1 - k_2 + k_3 - k_4)
		\delta( \prod_{i=1}^4 p_i - 1 ) \;
		E_{k_1p_1}^\dagger E_{k_2p_2}\zeta^\dagger_{k_3p_3}\zeta_{k_4p_4}
		\nonumber \\ \fl \hspace{1cm}
		\times p_1 \left( p_4 \gamma_{k_1-k_4}
			\cosh \varphi_{k_3p_3} \cosh \varphi_{k_4p_4}
		+ p_3 \gamma_{k_1-k_3}
			\sinh \varphi_{k_3p_3} \sinh \varphi_{k_4p_4} \right)
		\nonumber \\ \fl \hspace{1cm}
		- \frac{zt \cos 2\chi}{N} \sum_{k_1 \ldots k_4} \sum_{p_1 \ldots p_4}
		\delta^{(2)} (k_1 - k_2 + k_3 + k_4)
		\delta( \prod_{i=1}^4 p_i - 1 ) \;
		E_{k_1p_1}^\dagger E_{k_2p_2}\zeta^\dagger_{k_3p_3}\zeta^\dagger_{k_4p_4}
		\nonumber \\ \fl \hspace{1cm}
		\times p_1 \left( p_4 \gamma_{k_1+k_4}
			\cosh \varphi_{k_3p_3} \sinh \varphi_{k_4p_4}
		+ p_3 \gamma_{k_1+k_3}
			\sinh  \varphi_{k_3p_3} \cosh \varphi_{k_4p_4} \right) + h.c..
\end{eqnarray}
Finally, we can also write out the Hamiltonian for the interaction vertex with the transversal spin waves. We can write this Hamiltonian term explicitly using phase and momentum conservation,
\begin{eqnarray}
	\fl H^{T}_2 =
		- \frac{zt}{N} \sum_{k_1 \ldots k_4}\sum_{p_1 \ldots p_4}
		\delta^{(2)}( k_1 -k_2+k_3-k_4) \delta( \prod_{i=1}^4 p_i - 1 ) \;
		\; E^\dagger_{k_1 p_1} E_{k_2 p_2} 
	\nonumber \\ \fl \hspace{2cm}
		\times \left( \alpha^\dagger_{k_3 p_3} \alpha_{k_4 p_4}
			+ \beta^\dagger_{k_3 p_3} \beta_{k_4 p_4} \right)
	\nonumber \\ \fl \hspace{2cm}
		\times p_1 \left( p_4 \gamma_{k_1 - k_4} \cosh \theta_{k_3 p_3} \cosh \theta_{k_4 p_4}
		+ p_3 \gamma_{k_1 - k_3} \sinh \theta_{k_4 p_4} \sinh \theta_{k_3 p_3} \right)
	\nonumber \\ \fl \hspace{1cm}
		- \frac{zt}{N} \sum_{k_1 \ldots k_4}\sum_{p_1 \ldots p_4}
		\delta^{(2)}( k_1 -k_2+k_3+k_4)\delta( \prod_{i=1}^4 p_i - 1 ) \;
		\; E^\dagger_{k_1 p_1} E_{k_2 p_2} \alpha^\dagger_{k_3 p_3} \beta^\dagger_{k_4 p_4}
	\nonumber \\ \fl \hspace{2cm}
		\times p_1 \left( p_4 \gamma_{k_1 + k_4} \cosh \theta_{k_3 p_3} \sinh \theta_{k_4 p_4}
		+ p_3 \gamma_{k_1 + k_3} \cosh \theta_{k_4 p_4} \sinh \theta_{k_3 p_3} \right)
		+ h.c..
\end{eqnarray}

These expressions are used to transform the Feynman diagrammatic representation of the SCBA of Figure \ref{FigSCBA} into the explicit formula (\ref{SCBA}).

\end{document}